%% file: Linden2021_subm_pdf.tex
\documentclass[twocolumn]{aastex62}

\shorttitle{Massive Star Cluster Formation and Destruction in Luminous Infrared Galaxies in GOALS II}
\shortauthors{S. Linden et al.}

\begin{document}

\title{Massive Star Cluster Formation and Destruction in Luminous Infrared Galaxies in GOALS II: An ACS/WFC3 Survey of Nearby LIRGs}

\author{S. T. Linden}
\affiliation{Department of Astronomy, University of Massachusetts at Amherst, Amherst, MA 01003, USA.}

\author{A. S. Evans}
\affiliation{Astronomy Department, University of Virginia, 530 McCormick Road, Charlottesville, VA 22904 USA}
\affiliation{National Radio Astronomy Observatory, 520 Edgemont Road, Charlottesville, VA 22903 USA}

\author{K. Larson}
\affiliation{Space Telescope Science Institute, 3700 San Martin Drive, Baltimore, MD, 21218, USA}

\author{G. C. Privon}
\affiliation{National Radio Astronomy Observatory, 520 Edgemont Road, Charlottesville, VA 22903 USA}
\affiliation{Department of Astronomy, University of Florida, 211 Bryant Space Sciences Center, Gainesville, FL 32611}

\author{L. Armus}
\affiliation{IPAC, California Institute of Technology, MS 100-22, Pasadena, CA 91125 USA}

\author{J. Rich}
\affiliation{Carnegie Observatories, 813 Santa Barbara St., Pasadena, CA 91101}

\author{T. D\'iaz-Santos}
\affiliation{Department of Physics, University of Crete, GR-71003, Heraklion, Greece}

\author{E. J. Murphy}
\affiliation{National Radio Astronomy Observatory, 520 Edgemont Road, Charlottesville, VA 22903 USA}
\affiliation{Astronomy Department, University of Virginia, 530 McCormick Road, Charlottesville, VA 22904 USA}

\author{Y. Song}
\affiliation{Astronomy Department, University of Virginia, 530 McCormick Road, Charlottesville, VA 22904 USA}

\author{L. Barcos-Mu\~noz}
\affiliation{National Radio Astronomy Observatory, 520 Edgemont Road, Charlottesville, VA 22903 USA}
\affiliation{Astronomy Department, University of Virginia, 530 McCormick Road, Charlottesville, VA 22904 USA}

\author{J. Howell}
\affiliation{IPAC, California Institute of Technology, MS 100-22, Pasadena, CA 91125 USA}

\author{V. Charmandaris}
\affiliation{Department of Physics, University of Crete, GR-71003, Heraklion, Greece}
\affiliation{Institute of Astrophysics, FORTH, GR-71110 Heraklion, Greece}

\author{H. Inami} 
\affiliation{Hiroshima Astrophysical Science Center, Hiroshima University, 1-3-1 Kagamiyama, Higashi-Hiroshima, Hiroshima 739-8526, Japan}

\author{V. U}
\affiliation{Department of Physics and Astronomy, University of California, Irvine, 4129 Frederick Reines Hall, Irvine, CA 92697, USA}

\author{J. A. Surace}
\affiliation{Eureka Scientific, Inc. 2452 Delmer Street Suite 100 Oakland, CA 94602-3017 USA}

\author{J. M. Mazzarella}
\affiliation{IPAC, California Institute of Technology, MS 100-22, Pasadena, CA 91125 USA}

\author{D. Calzetti}
\affiliation{Department of Astronomy, University of Massachusetts at Amherst, Amherst, MA 01003, USA.}

\date{\today}

%BEGIN ABSTRACT~~~~~~~~~~~~~~~~~~~~~~~~~~~~~~~~~~~~~~~~~~~~~~~~~~~~~~~~~~~~~~~~~~~~~~~~~~~~~~~~~~~~~~~~~~~~~~~~~~~~~~
\begin{abstract}

%Relative to previous FUV-optical studies of LIRGs, the field of view and improved sensitivity of WFC3 provide over an order of magnitude increase in the total number of clusters detected in these systems. 
We present the results of a Hubble Space Telescope WFC3 near-UV and ACS/WFC optical study into the star cluster populations of 10 luminous and ultra-luminous infrared galaxies (U/LIRGs) in the Great Observatories All-Sky LIRG Survey (GOALS). Through integrated broadband photometry we have derived ages, masses, and extinctions for a total of 1027 star clusters in galaxies with $d_{L} <$ 110 Mpc in order to avoid issues related to cluster blending. The measured cluster age distribution slope of $dN/d\tau \propto \tau^{-0.5 +/- 0.2}$ is steeper than what has been observed in lower-luminosity star-forming galaxies. Further, differences in the slope of the observed cluster age distribution between inner- ($dN/d\tau \propto \tau^{-1.07 +/- 0.12}$) and outer-disk ($dN/d\tau \propto \tau^{-0.37 +/- 0.09}$) star clusters provides evidence of mass-dependent cluster destruction in the central regions of LIRGs driven primarily by the combined effect of strong tidal shocks and encounters with massive GMCs. Excluding the nuclear ring surrounding the Seyfert 1 nucleus in NGC 7469, the derived cluster mass function (CMF: $dN/dM \propto M^{\alpha}$) has marginal evidence for a truncation in the power-law (PL) at $M_{t} \sim 2$x$10^{6} M_{\odot}$ for our three most cluster-rich galaxies, which are all classified as early-stage mergers. Finally, we find evidence of a flattening of the CMF slope of $dN/dM \propto M^{-1.42 \pm 0.1}$ for clusters in late-stage mergers relative to early-stage ($\alpha = -1.65 \pm 0.02$), which we attribute to an increase in the formation of massive clusters over the course of the interaction.

%Thus, not only does rapid cluster disruption appear to be ubiquitous in LIRGs at all galactocentric radii, the magnitude of the differences between inner- and outer-disk cluster disruption are amplified relative to observations of nearby normal star-forming galaxies. Given this, it appears that the differences in the observed cluster mass function (CMF) observed for SSCs in our LIRG sample relative to normal star-forming galaxies is caused primarily by a rapid cluster disruption, which flattens the observed CMF relative to an underlying initial cluster mass function.

\end{abstract}

\keywords{galaxies: star clusters: general - galaxies: starburst - galaxies: ISM - infrared: galaxies}

%\maketitle

%BEGIN INTRODUCTION~~~~~~~~~~~~~~~~~~~~~~~~~~~~~~~~~~~~~~~~~~~~~~~~~~~~~~~~~~~~~~~~~~~~~~~~~~~~~~~~~~~~~~~~~~~~~~~~~~

\section{Introduction}

%By contrast, normal star-forming galaxies contain hundreds of smaller mass HII regions \citep{doc16}, suggesting a fundamental shift in the way stars form and evolve relative to galaxies in the early Universe.
Imaging surveys of high-redshift star-forming galaxies reveal that these systems tend to display turbulent disks, which host star-forming clumps with stellar masses of $\sim 10^{7-8} M_{\odot}$ and sizes of $0.5-5$ kpc \citep{bge04,dme09,ed10,livermore15}. These extreme clumps are $\sim100$x more massive than the typical giant molecular cloud (GMCs) observed in the local Universe, and give rise to super star clusters (SSCs) with masses which can reach $\geq 10^{6} M_{\odot}$; i.e., similar to the most massive known globular clusters and 1-2 orders of magnitude more massive than any young cluster or HII region found in nearby normal galaxies like the Milky Way (MW) or M51 \citep{kj99,bcm14,doc16,daa18}. 

While the total number of massive clusters formed is largely set by the gas density in the interstellar medium (ISM), high-resolution simulations of cluster formation show that strong stellar (internal) feedback not only changes the efficiency of cluster formation within GMCs, but also alters the dynamical state of the cluster and in some cases may rapidly disrupt the cloud altogether \citep[e.g.,][]{li17,grudic20}. Star clusters may also lose mass via interaction (external) with their galactic environment \citep[e.g.,][]{lamers05,gieles06}, resulting in low disruption rates in low pressure environments and high disruption rates in intensely star-forming regions of the ISM. From a theoretical perspective, after their formation clusters lose mass in three phases: (1) stellar evolution and feedback, (2) two-body relaxation-driven evaporation in their host galaxy's tidal field, and (3) tidal shocks resulting primarily from collisions with GMCs and passages of spiral arms \citep[see][]{krause20}. During the first phase, expulsion of natal gas by stellar winds and supernovae perturbs the clusters' potential and leaves up to 90\% of clusters unbound or destroyed \citep{ll03}. The latter stages of cluster evolution depend both on the clusters' location within the galaxy, and its overall density.

% with a trend for cluster disruption increasing in galaxies with increasing star formation rate surface density
However, there is no consensus for what the relative importance of both internal feedback and the external galactic environment (mass-dependent cluster disruption) have on altering the universality of star cluster properties \citep[see:][]{pz10,bcm10,nb12,fac12,fouesneau12,rc15,lcj17}. Observationally, the shape of both the cluster luminosity function (CLF) and cluster mass function (CMF) as well as the star cluster age distribution have been found to vary between quiescent and starburst galaxies in the local Universe \citep{gieles09, nb12, krumholz19, aa20}. While \citet{mf05} and \citet{bcm07} interpret the age distribution for star clusters in the Antennae galaxies (NGC 4038/39), the nearest example of a massive galaxy merger, as evidence for mass independent cluster disruption (e.g. feedback-driven gas dispersal) up to $\sim 10^{5} M_{\odot}$, \citet{lamers05} and \citet{esv14} have suggested that strongly varying tidal fields and collisions between GMCs of high surface density can be the dominant physical mechanism which disrupts clusters in such extreme environments. Indeed, simulations of equal-mass galaxy mergers \citep{dk11} find that 80-85\% of the cluster disruption seen may be accounted for by tidal shocks.

Luminous and Ultra-luminous infrared galaxies (LIRGs: defined as having IR luminosities $L_{IR} [8-1000\mu m\ > 10^{11} L_{\odot}$; ULIRGs: $L_{IR} > 10^{12} L_{\odot}$) host the most extreme stellar nurseries in the local Universe, with star-forming clump masses and radii similar to high-redshift galaxies \citep{klarson20}. The activity in LIRGs is largely interaction triggered, with the progenitors observed to be gas-rich disk galaxies involved in minor interactions (at the low luminosity end) or major merger events \citep{ss13a}. This framework is supported by simulations of gas-rich mergers which show the inward flow of star-forming gas via tidal and bar-driven dissipation to a nuclear starburst, and the eventual settling into early-type galaxies \citep{bah92,barnes04,hopkins06}. Further, these simulations demonstrate that the chaotic environment of the nuclear region is contrasted against the outer regions of a galaxy merger, where increases in the dense gas and star formation efficiency are comparatively mild \citep{moreno21}. Since local LIRGs can be studied at high-resolution across the entirety of their disks, they are ideal laboratories for testing location-dependent cluster formation and disruption in extreme merger-driven environments.

To better understand cluster formation and evolution in LIRGs, we previously combined our {\it Hubble Space Telescope (HST)}  F435W (B) and F814W (I) ACS/WFC data with far-UV ACS/SBC F140LP (FUV) data to age-date $\sim 400$ clusters in a combined sample of 22 LIRGs as part of the Great Observatories All Sky LIRG Survey (GOALS) \citep{goals,stl17}. Overall we found evidence of a steeper decline in the number of clusters per unit time as a function of increasing cluster age ($dN/d\tau \propto \tau^{-0.9}$ for $3 < \tau < 300$ Myr) relative to `normal' galaxies, which is a further indication that clusters in massive mergers suffer destruction at a much more rapid rate than normal star-forming galaxies.

However, our 2017 study was limited by three issues: {\it (i)} the small field of view (FOV) of the SBC (Solar Blind Channel) observations ($30^{\prime\prime} \times 30^{\prime\prime}$) often limited us to clusters in the central regions of the galaxies in our sample; {\it (ii)} the completeness-corrected cluster sample contained only the most massive clusters ($M \geq 10^{5} M_{\odot}$) in each galaxy due to the limited sensitivity achieved with the ACS/SBC; and {\it (iii)} age--extinction degeneracies were present for clusters within particular color ranges. Here, we present {\it HST} WFC3 U- (F336W) band observations of all 10 $L_{\rm IR} > 10^{11.4} L_{\odot}$ U/LIRGs in GOALS scheduled to be observed by the {\it James Webb Space Telescope (JWST)} as part of the guaranteed time observing (GTO) campaigns and the early release science (ERS) program {\it ``A JWST Study of the Starburst-AGN Connection in Merging LIRGs''} (PID 1328 - PI Armus). The combination of broad- and narrow-band near-infrared (NIR) and mid-infrared (MIR) imaging with JWST at comparable resolution as HST will allow us to identify all of the very youngest ($\sim 1$ Myr) and most highly-embedded SSCs in LIRGs. These sources are often missed in UV-optical studies of cluster formation in nearby galaxies, but are crucial for a complete picture of stellar feedback in the ISM.

Crucially, our new WFC3/UVIS observations provide vast improvements in sensitivity over prior {\it HST} detectors operating at 0.3 $\mu{\rm m}$, and combined with our existing {\it HST} data (ACS/WFC F435W and F814W data), provide high-resolution maps of the distribution of star clusters over the entire extent of each LIRG. Further, the F336W filter provides a much improved lever arm for age-dating clusters relative to the F140LP filter, which allows us to break the reddening--age degeneracy and thus more accurately measure (to within 0.1-0.2 dex in $\log(\tau)$) cluster ages and cluster masses. A fundamental goal of the present paper is to investigate what role the host galaxy environment, and {\it localized} ISM conditions within individual LIRGs, play in the rapid destruction of SSCs observed in LIRGs.

The paper is organized as follows: In \S 2, the observations, data reduction, and sample selection are described. Our method for identifying clusters and determining cluster ages, masses, and extinctions is described in \S 3. In \S 4 we discuss the completeness limits of our cluster sample. In \S 5, the age and mass functions are discussed within the context of lower luminosity star-forming galaxies and clusters found in both the inner and outer-disks of our galaxy sample. Finally, we discuss our results in terms of the merger stages (MS) of each system. \S 6 is a summary of the results. 

Throughout this paper, we adopt a WMAP Cosmology of $H_0 = 69.3$ km s$^{-1}$ Mpc$^{-1}$, $\Omega _{\rm matter} = 0.286$, and $\Omega _{\Lambda} = 0.714$ \citep[e.g., see][]{wmap}.

\section{Observations and Sample Selection}

\input{tbl-1_full_v2}

\begin{figure*}
  \centering
  \includegraphics[scale=0.4]{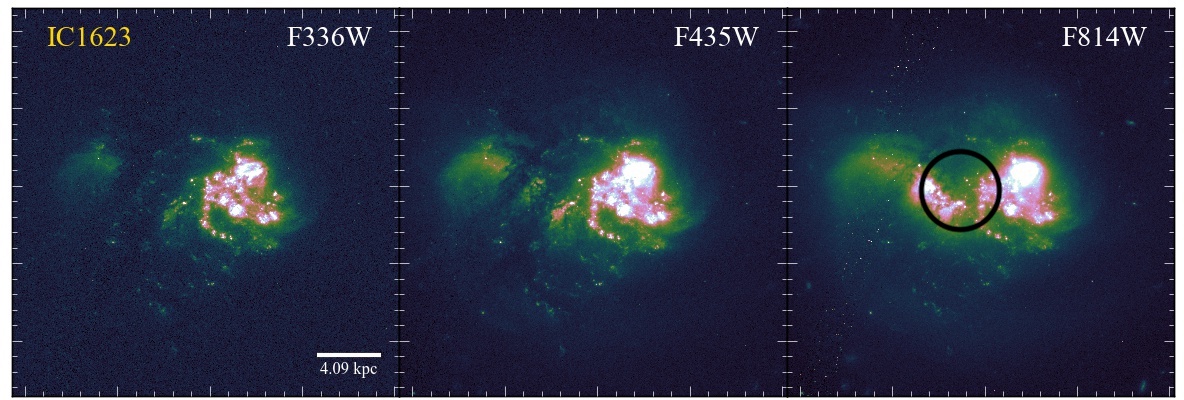}
  \caption{A comparison of {\it HST} WFC3/UVIS F336W image of IC 1623, relative to archival ACS/WFC F814W and F435W images. All three images have a FOV of 30" and are oriented N up E left. A corresponding 10" scale bar is given in the lower-right of the left Panel. The black circle in the right Panel represents the MIR size reported in Table 1 and discussed in Section 3.1. These images demonstrate our ability to detect and characterize star clusters throughout the central regions, disks, and extended tidal features of the galaxies in our sample.}
\end{figure*}

The F336W images of 10 LIRG systems were obtained from August - November 2018 (PI: A. Evans; PID 15472). Table 1 is a summary of the general properties of the sample. Each galaxy was observed with four dithered exposures in ACCUM mode, and when possible, placed at the center of the UVIS2 chip, due to the increased sensitivity at UV wavelengths relative to UVIS1. The approximate integration times for each galaxy is 41 minutes. The data were reduced with the Multidrizzle software provided by STScI to identify and reject cosmic rays and bad pixels, remove geometric distortion, and finally, combine the resulting mosaics with our existing ACS/WFC F435W and F814W observations (see Figure 1). The final pixel scale of our WFC3 images is $0.0396"/$pix. The ACS/WFC and WFC3/UVIS cameras have comparable fields of view (202"x202" and 162"x162" respectively), and cover sufficient area to enable single pointing observations of these U/LIRGs.

To understand the impact of combining results from galaxies which span a factor of $\sim3$ in distance, we performed simulations of cluster blending: by progressively smoothing {\it HST} WFC3 archival imaging of the Antennae Galaxies ($D_{\rm L} = 22$ Mpc: \citet{bcm10}, re-identifying star-clusters, and re-computing the luminosity function, we determined that blending begins to hamper our ability to accurately measure cluster properties (age, mass, extinctions), and thus determine the true slope, $\alpha$, of the underlying cluster mass function, at smoothing lengths of $\geq$ 5 pixels (i.e., $\Delta \alpha > 0.3$). This test demonstrates that clusters can be individually detected and their physical properties can be accurately recovered in galaxies out to a distance of $\sim 100$ Mpc. Therefore we limit our primary cluster analysis in this paper to the 5 nearest
systems that have have $D_{L} \leq 100$ Mpc, such that we can be confident our results are not biased towards large, unresolved, complexes of multiple star clusters. However, we run the same cluster identification and photometric analysis
on all 10 LIRG systems. We provide photometry and derived physical quantities for all star clusters in the Appendix.

\section{Cluster Identification and Model Fitting} 

\subsection{Cluster Identification}

\input{tbl-2}

Star clusters in all three bands were selected using the program Source Extractor \citep{sex}. The identification of clusters and the extraction of photometry is performed after a diffuse background subtraction is applied to each image. This subtraction is described in detail in \citet{stl17}, and is designed to remove starlight which is unassociated with the compact sources identified in each galaxy. Cluster photometry across all background-subtracted images was then calculated using the IDL package APER. We used an aperture of radius 3.0 pixels, with an annulus from 4-5 pixels to measure the local background surrounding each cluster. Aperture corrections for ACS/WFC and WFC3 were calculated based on the encircled energy values presented in \citet{sirianni05} and \citet{deustua16} respectively. We additionally applied a correction for foreground Galactic extinction, using the \citet{schlafly11} dust model combined with the empirical reddening law of \citet{fitz99} available through the NASA Extragalactic Database (NED). We also removed sources with a signal-to-noise ratio S/N$< 3$ and which are not detected in all three filters. 

We extracted photometry for 2167 cluster candidates ($N_{det}$) which is given for each galaxy in Table 2. We then used ISHAPE \citep{larsen99} to measure the 2-D FWHM for all remaining sources by de-convolving the {\it HST} instrumental point spread function with a King profile \citep{king66}. We find that a cut of 3 pixels FWHM (corresponding to an average cluster size of $R_{eff} \sim 22$~pc) effectively removes extended sources in both the nearest and farthest galaxies in the sample. This value is consistent with the upper-end of sizes observed for open star clusters in the Milky Way, as well as simulations of star cluster formation in dwarf starburst galaxies \citep{lahen20}. In Figure 2 we show a zoom-in for three regions in IC 1623 with our cluster detections overlaid in blue squares. These regions highlight the structure and variable levels of background emission which surround these clusters. A total of 1027 confirmed clusters across the 5 nearest LIRGs meet the above criteria. We observe a large range of cluster magnitudes ($M_{U} = -7 \sim -17$ mag), with a median of $M_{U} = -10$ mag, where the brightest clusters are detected in both the central region and outer-disks of these systems. The number of confirmed clusters for each galaxy are given in Table 2.

Finally, in order to make detailed comparisons of clusters at different radii we adopt a similar methodology to \citet{stl19}, to compute the de-projected galactocentric radius of each cluster ($r_{G}$) using the axis-ratio ($b/a$) and position angle (PA) of the host galaxy \citep{tj00,hyleda,dck13}. The $13.2\mu$m mid-infrared size derived from Spitzer/IRS spectra serves as our best estimate (or upper limit) for the size of the central starburst, limited by the resolution of the Spitzer/IRS beam, within each galaxy \citep{tds10}. By comparing the measured $r_{G}$ values to the $13.2\mu$m FWHM determined from a gaussian fit to the 2D spectra, we can place clusters into two categories: those that fall within the MIR size of the galaxy (inner-disk clusters) and those that fall outside of the MIR size (outer-disk clusters). This separation allows us to test whether the cluster properties that we measure with HST are dependent on the current size of the starburst responsible for producing the bulk of the far-infrared emission. 

We note that for several galaxies the MIR size is unresolved (FWHM is within 10\% of that of the unresolved stellar PSF). Therefore while the $13.2\mu$m MIR size may not represent the true size of the compact starburst in each galaxy, it still serves as our best estimate for where the bulk of the SF activity responsible for heating the dust in each galaxy is. We also stress that for NGC 3256 and IC 1623 (the two galaxies we ultimately examine - See Section 5.1.1) the MIR sizes are resolved. In fact for high-resolution ground-based $5-15\mu$m imaging of VV 114  reveal that up to $\sim 60\%$ of the MIR emission is extended on kpc-scales \citep{el02}. The values for the $b/a$, PA, and MIR size (FWHM) of each system are given in Table 1.

\subsection{Model Fitting}

For all confirmed clusters, the measured 3-band fluxes were compared against a library of simple stellar population (SSP) evolutionary models that were computed using the isochrone synthesis code Starburst99 \citep{sb99}, hereafter referred to as SB99. This code computes the evolution of an instantaneous burst based on a Kroupa IMF and the stellar evolution models over an age range of 1 Myr to 10 Gyr \citep[e.g.,][]{padova94}. Crucially we add the contribution from nebular emission lines which may contribute up to 40\% of the emission observed in the broad-band F814W filter for clusters with ages $\sim 3-5$ Myr \citep{reines09,yggdrasil}. We also choose to adopt a solar metallicity, as suggested for LIRGs by \citet{kewley10}, and the starburst attenuation curve of \citet{dc00}. Although we only consider solar metallicity models in our analysis, \citet{stl17} demonstrated that including sub-solar metallicity models does not significantly impact the results. The total attenuation of the stellar continuum, $R_{V} = A(V)/E(B-V)_{*} = 4.05 \pm 0.8$, is calibrated specifically for starburst galaxies and differs from the typical Milky Way value of $R_{V} \sim 3.1$ \citep{dc00}. It has been shown empirically that clusters and HII regions are more heavily attenuated than the underlying stellar continuum, due to the fact that these objects are often found near dusty regions of ongoing star formation \citep{dc94}. Our model choices allow us to make consistent comparisons with results from the Legacy Extragalactic UV Survey (LEGUS), which adopt the Yggdrasil nebular+continuum models \citet{yggdrasil} to determine cluster ages, masses, and extinctions \citep[e.g.,][]{ryon17,aa17,doc19}.

\begin{figure*}
  \centering
  \includegraphics[scale=0.6]{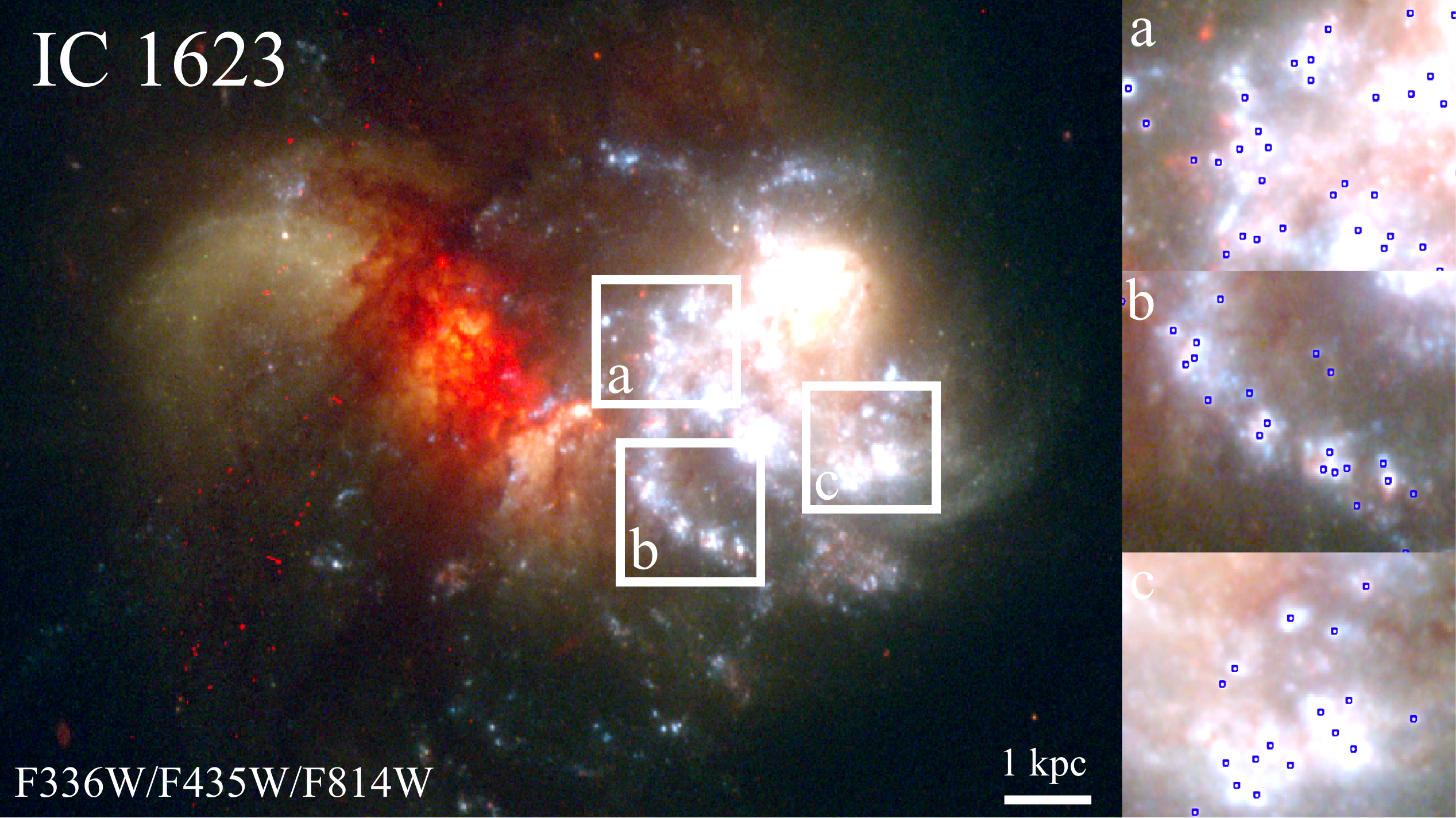}
  \caption{A false-color image of IC 1623 constructed with the HST WFC3 F336W and the ACS/WFC F435W and F814W filters in the blue, green, red channels respectively. The 3 regions labeled a, b, c have cluster detections identified in the panels at right. In these panels we overlay cluster detections in blue circles. These cutouts demonstrate that we are able to detect UV-bright compact sources of emission throughout the disk of the eastern galaxy, where the local background shows significant variations and structure. Approximately 200 star clusters are detected.}
\end{figure*}

We construct grids for the age, mass, and extinction model parameters and perform a $\chi^{2}$-fit for each point in the grid. We use these to obtain the marginalized posterior distribution function for all derived parameters, and represent the error in our model fit as the FWHM of each distribution. In the Appendix we give an example of our best-fit model one cluster in NGC 3256, chosen to demonstrate our ability to accurately recover the properties of a young, moderately extincted ($A_{V} = 1.7$), massive ($10^{4} M_{\odot}$) cluster. We show both the maximum a posteriori probability (MAP: blue line) as well as the median (pink line) of each parameter along with the corresponding $1\sigma$ errors. This cluster represents a prototypical degenerate case from our previous F140LP observations, and with the substitution of the F336W filter we are able to recover cluster age, mass, and extinction to within 0.2 dex. This is crucial for accurately recovering the underling slope of the age and mass distribution functions for each galaxy. The median age, mass, and extinction along with median absolute deviation (MAD) of each distribution are given in Table 2.
%it is clear that

\section{Completeness and the Mass-Age Diagram}

\subsection{Completeness}

In order to determine the completeness limit of our cluster sample, we follow a similar prescription to \citet{mm18} which utilizes the LEGUS Cluster Completeness Tool. We first create a set of synthetic clusters with the BAOlab software \citep{larsen99} by convolving the WFC3 F336W PSF with a King profile specified with effective radii of 1, 5, and 10 pc. The resulting sources are then distributed within our original science frames in two ways: (1) 100 clusters with apparent magnitudes of $18 < m_{U} < 26$ are randomly placed within the central 300 pixels, which is chosen to match the median MIR size in our 5 sample galaxies, of each F336W science frame, and (2) 500 clusters are randomly placed outside the central 300 pixels of each science frame. In this way we can directly account for significant variations in the depth of our U-band images in the central regions of the LIRGs in our sample. The same procedure described in \S 3 is then applied to these images and the fraction of simulated clusters recovered in the final catalog is calculated for each effective radius as a function of input cluster magnitude.

For the 5 LIRGs in our cluster sample the total $90\%$ completeness limits for the inner and outer-region are given in Table 2. We find that due to increased dust obscuration and crowding in the central regions we are $\sim 1$ magnitude less sensitive in the F336W filter, and that we achieve consistent depth (m$_{U, Inner} 90\%$) across the 5 galaxies in our sample given that the range in magnitudes is relatively small. The cluster-weighted median completeness limit for the inner and outer regions of our LIRG sample are $M_{U} = -9.3$ and $M_{U} = -8.4$ respectively, and are shown as dotted and solid lines in Figure 3. 

\subsection{The Mass-Age Diagram}

Figure 3 shows the derived age and mass of each cluster identified in the sample with outer-disk clusters ($N_{c} = 809$) plotted in black and inner-disk clusters ($N_{c} = 218$) plotted in purple. The lack of low-mass ($\leq 10^{4} M_{\odot}$) clusters with ages $\geq 100$ Myr is due to the fact that clusters dim as they age and eventually become fainter than our UV detection limits.  By applying our inner-disk completeness limit to the SB99 model we can define regimes of this mass-age parameter space where we are observationally complete in both the inner- and outer-disks of the LIRGs in our sample. Mass-limited cluster samples have the advantage over luminosity-limited samples because they recover the underlying shape of the age and mass distributions, and are thus not affected by the distance to each galaxy \citep{renaud20}.

%We emphasize that this $50\%$ limit for the sample is not a strong function of the distance to any galaxy.
%We also note the large number of clusters seen with ages below 10 Myr over the full range of masses. Although the cluster fitting method can create some observed structure in the mass-age diagram, it is unlikely to do so over all masses at young ages. In particular the lack of clusters with ages of $\sim 15$ Myr is a common feature of model-derived mass-age diagrams of star clusters in galaxies \citep{gieles05,goddard10}.

\begin{figure*}
\centering
\includegraphics[scale=0.65]{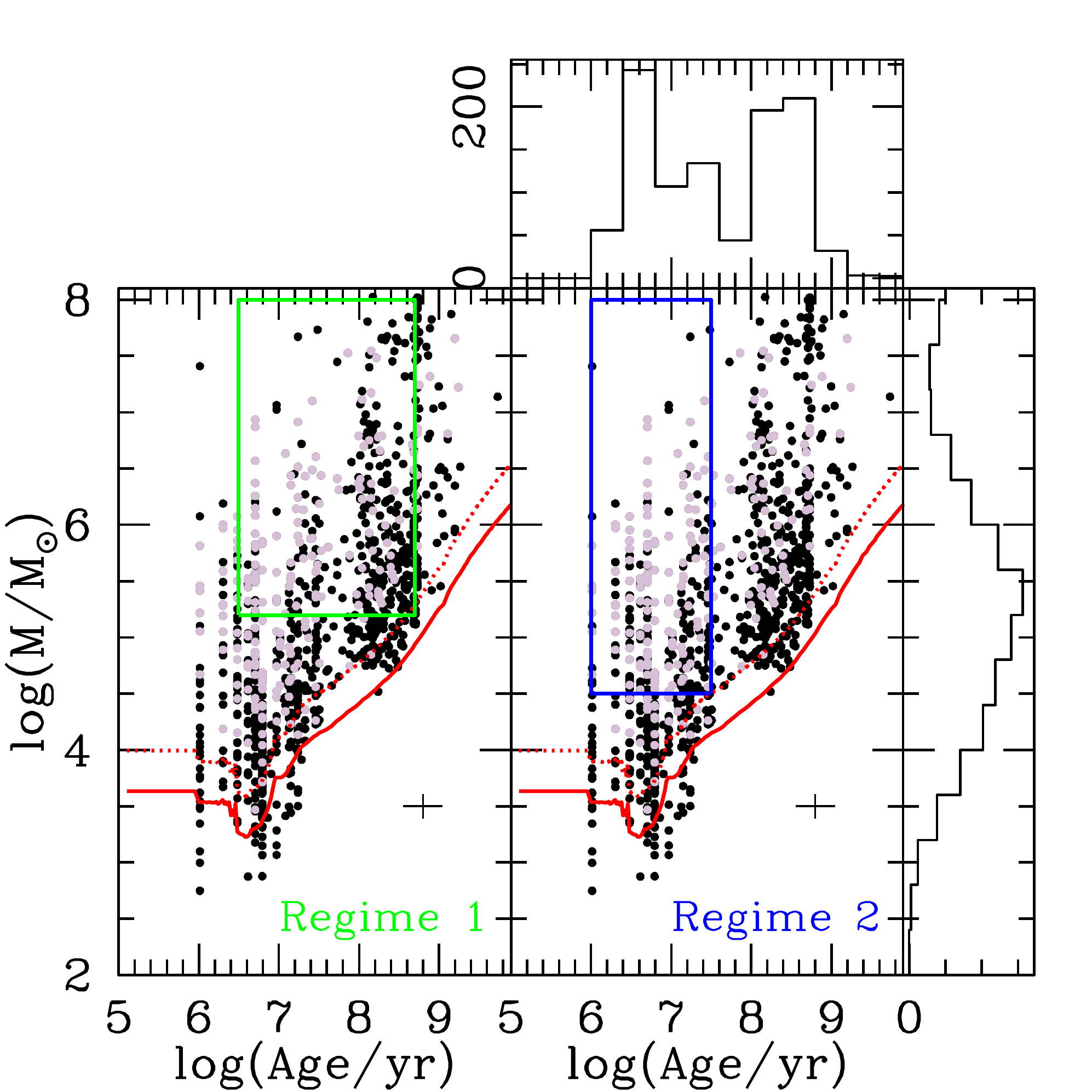}
\caption{The mass, age distribution of all 1027 clusters found in our LIRG sample. Outer- and inner-disk clusters, categorized based on the MIR size of the galaxy, are plotted in black and purple respectively. The solid and dotted red curves represent mass-age tracks produced from the SB99 model with an input of $M_{U} = -8.4$ and $M_{U} = -9.4$ which correspond to the $90\%$ completeness limits for the inner- and outer-disk respectively. The green box in the left panel represents Regime 1, and is used for the mass-age cuts applied when analyzing the cluster age distribution. The blue box in the right panel represents Regime 2, and is used for the mass-ages cuts applied when analyzing the cluster mass distribution. The histograms show the distribution of cluster ages and masses for the full sample. The cross on the bottom right of each panel represents the median errors in cluster age and mass bootstrapped from our model.}
\end{figure*}

The two cuts were selected to sample distinct parts of the mass and age distribution for which we maintain completeness. We define Regime 1 to be:

\begin{equation}
  5.2 < log(M/M_{\odot}) < 8
\end{equation}
\begin{equation}
  6.6 < log(\tau) < 8.7
\end{equation}

\noindent
Regime 2 to be:

\begin{equation}
  4.5 < log(M/M_{\odot}) < 8
\end{equation}
\begin{equation}
  6.0 < log(\tau) < 7.5
\end{equation}

\noindent
Regime 1 is chosen to match, as closely as possible, the age and mass limits from \citet{mf05,stl17}, allowing us to make accurate comparisons to the age distributions of clusters in other nearby galaxies, while also avoiding selecting the very youngest clusters which may actually be unresolved unbound associations \citep{kab16}. Regime 2 is chosen in order to measure the CMF of the young ($\tau \leq 10^{7.5}$), recently formed clusters that best reflect the current conditions of the ISM in these systems \citep{aa20}. When analyzing Regimes 1 and 2 we exclude the largest mass bin of $log(M/M_{\odot}) = 8.0$. These very high masses are most likely the result of either an imperfect extinction correction or multiple very compact star clusters in close proximity appearing as a single star cluster at the resolution of these images. We stress that the adopted FWHM cut helps minimize the inclusion of sources which are blends of $\sim 2-3$ clusters. Finally, we note that while clusters of these masses are rare, \citet{nb13b} found several clusters in NGC 7252 with masses greater than $10^{7} M_{\odot}$, including one cluster with a total mass of $\sim 10^{8} M_{\odot}$. Finally, recent observations of a nearby ($z=0.0336$) post-starburst galaxy have revealed massive clusters up to $10^{7.5} M_{\odot}$ associated with a period of intense SF activity in the galaxy \citep{chandar21}.

%BEGIN DISCUSSION SECTION~~~~~~~~~~~~~~~~~~~~~~~~~~~~~~~
\section{Discussion}

After determining ages, masses, and extinctions for the entire cluster sample we directly compare these distributions with those of nearby normal and interacting galaxies. We focus on the interpretation of the derived cluster age distribution for both the individual systems and the sample as a whole as well as mass functions for combined samples of galaxies. Further, we will analyze the inner- and outer-cluster distribution functions for each system. Ultimately, we discuss to what degree the differences observed in our cluster populations can be attributed to the extreme star-forming environment unique to LIRGs in the local Universe.

\subsection{Age Distribution}

\begin{figure}
\centering
\includegraphics[scale=0.4]{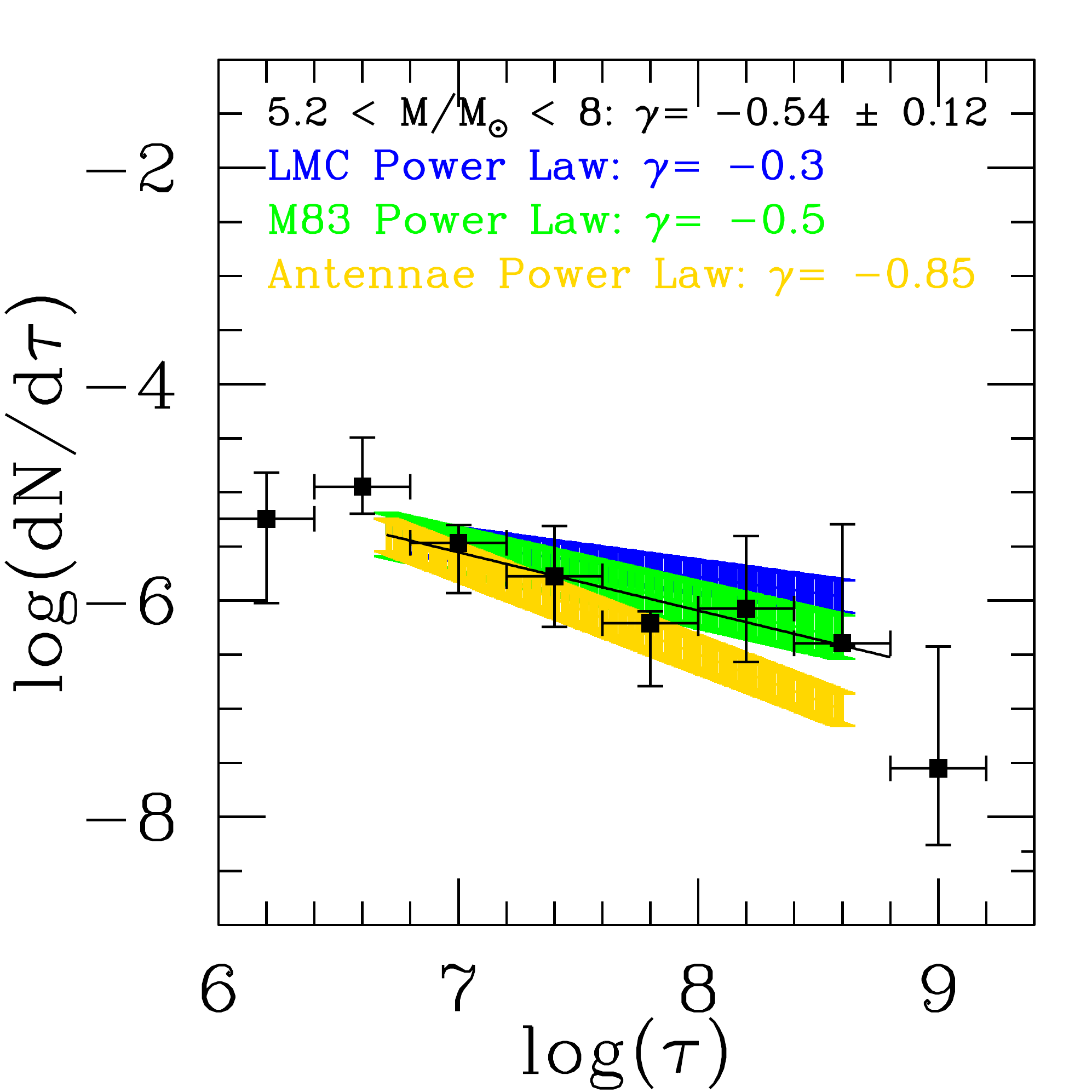}
\caption{The stacked age distribution functions for all 5 galaxies. The black line represent weighted linear least squares fit to the data in Regime 1. The blue, green, and yellow age functions of the LMC, M83, and the Antennae respectively, chosen to span the full range of observed cluster disruption rates, are taken from \citet{aab15} and normalized to the total number of clusters in our sample. It is clear LIRGs have elevated levels of cluster disruption relative to lower-luminosity star-forming galaxies like the LMC.}
\end{figure}

We consider the age distribution of clusters in our LIRG sample over Regime 1 described in \S 4. Specifically, we are interested in measuring the power law index $\gamma$, where $dN/d\tau \propto \tau^{\gamma}$. In Figure 4 we show the differential number of clusters per time interval, $\log (dN/d\tau)$, versus the median cluster age over that interval, $\log (\tau)$ for all clusters in the mass range given in Equation 1. The data are binned by 0.4 in $\log (\tau)$ so as to fully encapsulate $\sim 2\sigma$ times the typical model errors of 0.2 in $\log (\tau)$ discussed in \S 3. For clusters in the age-range given in Equation 2 a weighted linear least-squares fit to the age distribution results a power-law index of $\gamma = -0.54 \pm 0.12$. The slope of the cluster age distribution reflects a combination of any increase in the formation rate of clusters with $10^{5.2} M_{\odot} < M_{c} < 10^{8} M_{\odot}$ from $10^{8.7}$ yr to the present day as well as any destruction of star clusters through the methods discussed above over the same age interval. In Section 5.1.2 we attempt to account for this ambiguity directly.

While the observed slope is slightly shallower than both the Antennae age distribution slope and the results from \citet{stl17}, we see that our combined galaxy slope is consistent with the results for M83 and steeper than what is measured for lower-luminosity star-forming galaxies like the LMC \citep{aab15}. When examining the age distribution functions for our three most cluster-rich (i.e. $N_{c} > 100$ SSCs) galaxies, we find that the slopes measured for IC 1623, NGC 3256, and the outer-disk of NGC 7469 are all steep with $\gamma = -0.4$ to -1 (Figures 5, 6, and 7 respectively). 

We note that this analysis excludes the 19 clusters identified in the nuclear ring of NGC 7469. The star and cluster formation within nuclear rings is likely decoupled from the SFH of their host galaxies due to irregularities in the inflow rate of dense gas \citep{maoz01}. Indeed, $\sim 30\%$ of the nuclear K-band light in NGC 7469 is due to a single source \citep{davies04}. Further NGC 7469 is the only nearby galaxy in our sample that contains a bright Seyfert 1 nucleus, which may also affect the temperature and spatial distribution of the gas within the ring itself \citep{td07}. Therefore, we are unable to disentangle these effects from variations in the SFH which ultimately complicates any comparison between the age distribution inferred for these clusters the outer-disk clusters in this galaxy. In the following Section we exclude this galaxy from the inner- and outer-disk comparison.

\begin{figure}
\centering
\includegraphics[scale=0.4]{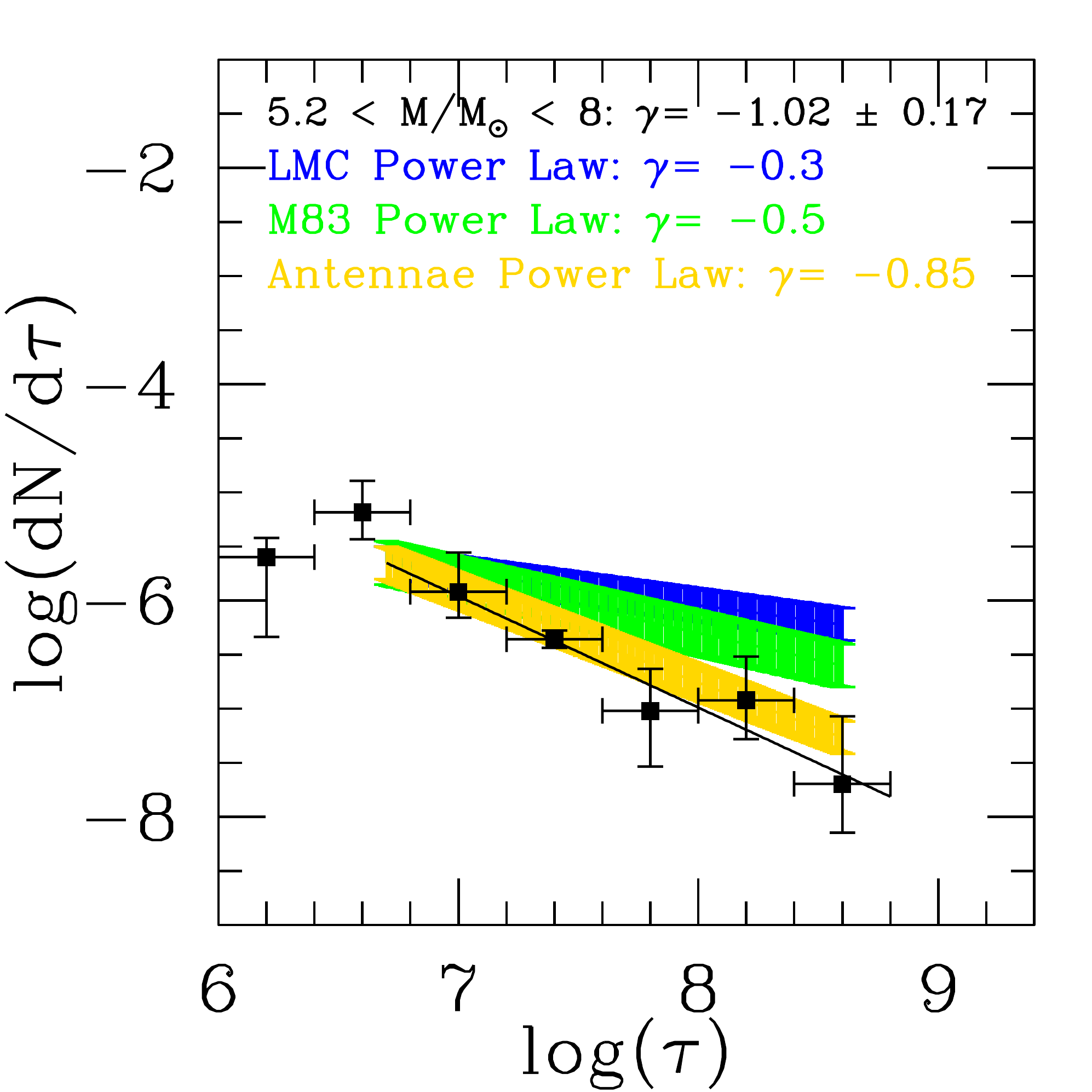}
\caption{The age distribution function for IC 1623 which contains 180 confirmed SSCs. We find that this galaxy shows the steepest slope ($\gamma = -1.02 \pm 0.17$) in our sample, and is steeper than the slopes observed in both M83 or the LMC. This value is consistent with the results presented in \citet{stl17} as well as observations of the Antennae galaxy \citep{mf05}.}
\end{figure}

\begin{figure}
\centering
\includegraphics[scale=0.4]{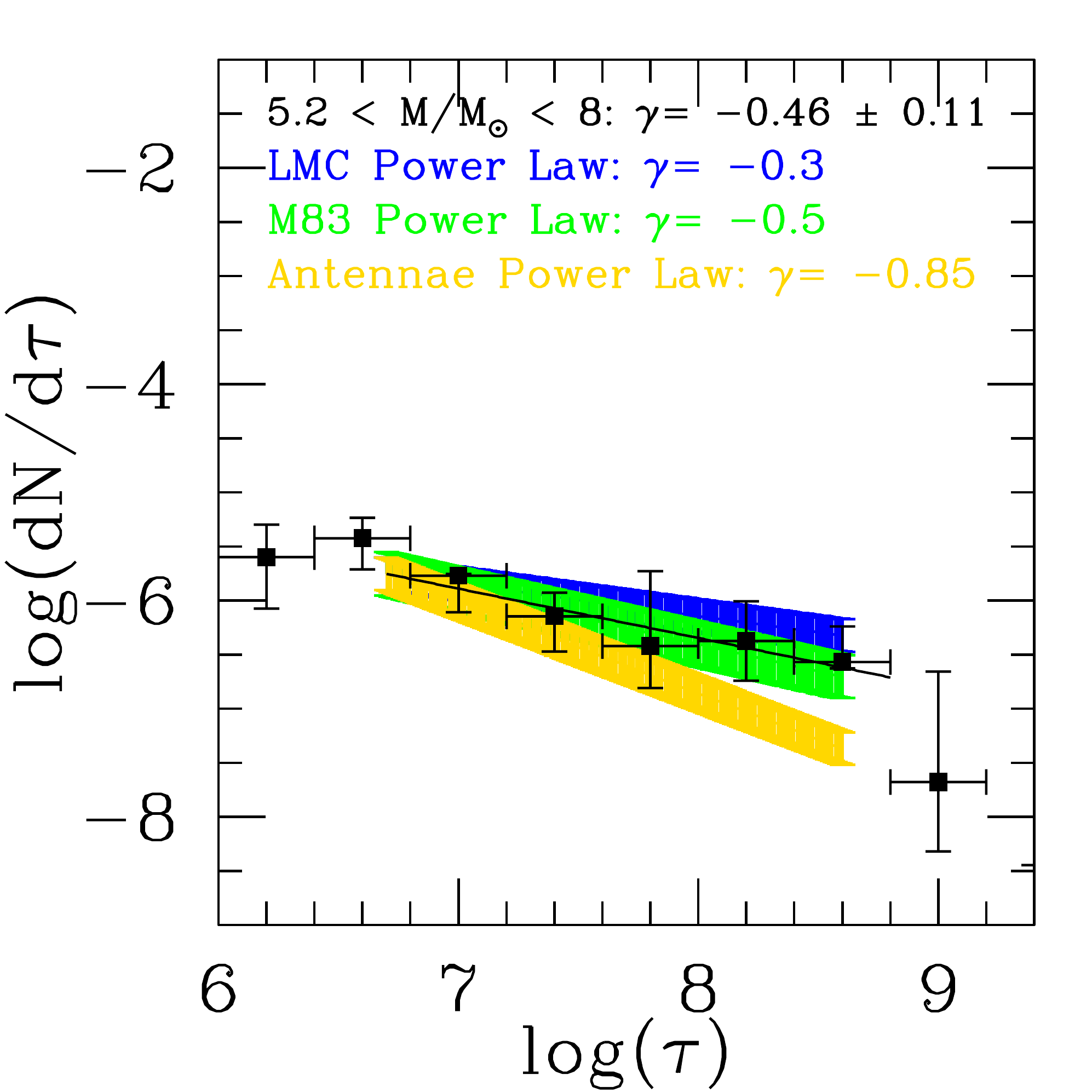}
\caption{The age distribution function for NGC 3256 which contains 549 confirmed SSCs. We find that the distribution slope ($\gamma = -0.46 \pm 0.11$) is consistent with the overall slope measured for the full sample, and importantly, is found to be flatter than what is measured for the Antennae galaxy \citep{mf05}.}
\end{figure}

\begin{figure}
\centering
\includegraphics[scale=0.4]{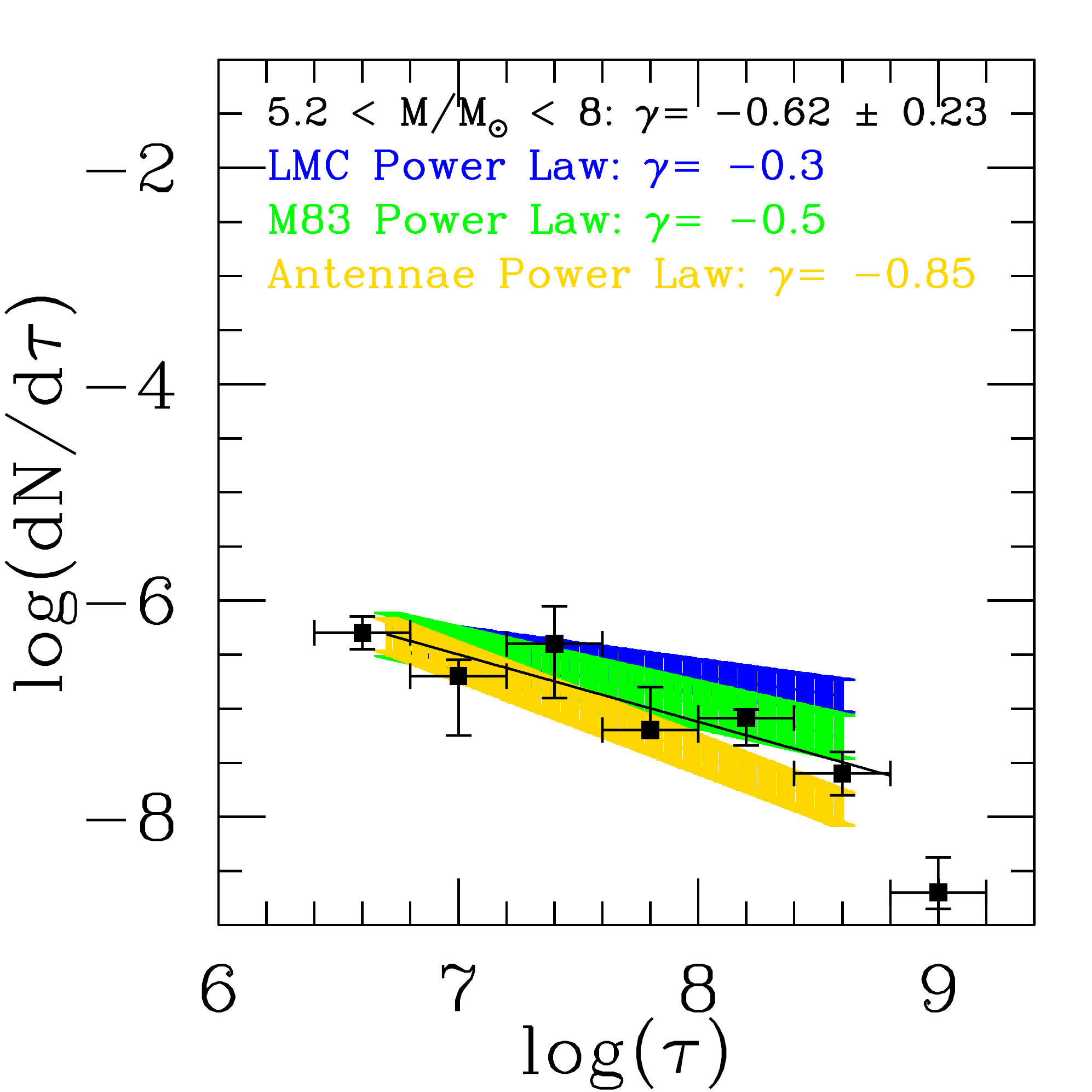}
\caption{The age distribution function for NGC 7469 which contains 146 confirmed SSCs. After excluding clusters in the nuclear ring, we find that the distribution is consistent with the overall slope measured for the full sample, and importantly, is found to be steeper than what is measured for lower-luminosity star-forming galaxies like the LMC.}
\end{figure}

\subsubsection{Inner- and Outer-disk Clusters}

All three 'cluster-rich' systems are classified as early-stage mergers \citep{dck13}, suggesting that the similar dynamical state (e.g tidal field) within the ISM of these galaxies may set the overall disruption rate seen. Existing wide-field WiFeS IFU observations of IC 1623 and NGC 3256  allow us to connect the cluster properties in both the inner- and outer-disks of these systems to the ionized gas. These two systems both exhibit strong evidence of galaxy-wide shock excitation induced by their ongoing merger activity \citep{jrich11}. Multi-line Atacama Large Millimeter/submillimeter (ALMA) CO observations of GMCs in IC 1623 and NGC 3256 also suggest they are highly-turbulent \citep{saito15,brunetti21}. With multi-line HCN/HCO+ observations \citet{saito18} further attributing this turbulence to galaxy-wide shocks in IC 1623.

By examining the age distribution functions for clusters identified as inner- or outer-disk clusters in NGC 3256 and IC 1623 a striking result emerges. The SSCs classified as inner-disk ($N_{c} = 206$) show a distribution slope of $\gamma = -1.1 \pm 0.11$, whereas SSCs found in the outer-disk ($N_{c} = 523$) have a distribution slope of $\gamma = -0.33 \pm 0.12$ (Figure 8). This inner-disk value is consistent with the results from \citet{stl17}, where the FOV is well-matched to the average MIR size. We stress that NGC 3256 and IC 1623 both individually show increases in the distribution slope between inner and outer-disk clusters, but the samples are combined here such that we achieve better statistics for each region. The outer-disk value is consistent with the results for M83 presented in \citet{aab15}, which is a prototypical spiral galaxy in the local Universe, and suggests that differences in the observed slope between IC 1623 and M83 in Figure 5 are driven by the inner region of the galaxy.

It was shown in \citet{gieles06} that star clusters can be disrupted on $\sim1-2$ Gyr timescales by tidal shocks that arise from gravitational interactions with passing GMCs. The GMC volume density in the central regions of LIRGs is considerably higher than what is seen in other nearby galaxies or the MW \citep[e.g,][]{rosolowski05,hughes13,greave09,papadopoulos12}. Therefore we might expect more frequent and/or more disruptive shocks, capable of disrupting even the most massive clusters on timescales of $\leq 100$ Myr, to be prevalent in the youngest and densest star-forming regions of local LIRGs \citep[often referred to as the 'cruel cradle effect':][]{bge10}. For very short disruption timescales, this scenario predicts that cluster evolution may be strongly dependent on environment \citep{miholics17}.

It is likely that both internal feedback and external effects co-exist for the youngest clusters. However, there's a suggestion from simulations that the fraction of clusters which are destroyed by gas expulsion \citep[i.e., 'infant mortality':][]{ll03} decreases in effectiveness with increasing ambient density of the star-forming region, while at the same time the disruptive effect of tidal shocks increases with ambient gas density \citep{pfeffer17}. It may be that 'infant mortality' and the 'cruel cradle effect' each dominate as the main feedback mechanism at the two extremes of the gas density spectrum of star-forming regions. Their relative strength would then determine the relation between the cluster formation efficiency (CFE = CFR/SFR) and the surface density of gas in galaxies \citep{lcj17, aa20}.

Assuming spatially homogeneous distributions of clusters and GMCs, \citet{gieles06} found that clusters lose most of their mass in very close encounters with high relative velocities. The derived disruption timescale depends on the cluster mass ($M_{c}$), the half-mass radius ($r_{h}$), and the product of the individual mean GMC surface density and the average local GMC density ($S \propto (\Sigma_{GMC} (M_{\odot} pc^{-2})\rho_{GMC} (M_{\odot} pc^{-3}))^{-1}$).

In Equation 5 we give the expression for the disruption timescale $t_{dis}$ for a cluster embedded in a strong external tidal field S parameterized by the mass density and volume of the ISM in the Solar Neighborhood \citep[Eq. 24 - ][]{gieles06}:

\begin{equation}
 t_{dis} = 2.0 \left(\frac{5.1 M_{\odot}^{2} pc^{-5}}{\Sigma_{ISM} \rho_{ISM}}\right) \left(\frac{M_{c}}{10^{4} M_{\odot}}\right)^{\delta} \textrm{Gyr}
 \end{equation}

where $\delta = 1 - 3\lambda$ with $R \propto M^{\lambda}$. Simulations of the mass-radius relationship suggest that the appropriate value of $\delta$ ranges from 0.2 for low-mass clusters to $\sim 1$ for the highest-mass (weakly-disrupted) clusters \citep{gr16}. In the latter case corrections due to adiabatic compression become so important that tidal shocks can destroy clusters in a single encounter \citep{miholics17}. Observations of nearby galaxies including M51, M33, and the SMC suggest a weak mass-radius relation \citep[$\lambda \sim 0.1$ resulting in $\delta = 0.6 - 0.7$:][]{boutloukos03}. However, recent results from LEGUS show that the average $\lambda$ is 0.24, although the median cluster mass in the LEGUS sample is 1-2 orders of magnitude lower than the clusters studied here \citep{bag21}.
 
Using observations of the molecular gas scale height ($H = 177$ pc), and GMC surface density ($\Sigma = 2.5$x$10^{3} M_{\odot}$pc$^{-2}$) in the central regions of NGC 3256, as well as an average cluster mass of $10^{5.8} M_{\odot}$, we derive a disruption timescale of $\sim 10$ Myr \citep{wilson19,brunetti21} which is comparable to the timescale over which we expect internal feedback processes to occur ($\sim 3-5$ Myr). The upper-limit represents the point where massive stars begin to die in supernova explosions injecting energy and momentum into their surroundings \citep{chevance20a, barnes20}. This is contrasted against a disruption timescale for the outer-disk of $\sim 200-300$ Myr, with an order of magnitude lower ISM density and approximately the same median cluster mass ($10^{5.8} M_{\odot}$). 

\citet{gr16} presents an updated version of Equation 5 which additionally accounts for internal evolution of the cluster during this disruption phase as well as the velocity dispersion of the surrounding gas. Depending on the adopted mass-radius relationship this results in an increase in the derived disruption timescale by a factor of $\sim 3-30$. For the central region of NGC 3256 we derive a disruption timescale of $\sim 40$ Myr for the inner-disk and $\sim 400$ Myr of the outer-disk adopting values for the velocity dispersion of the gas of 120 km/s and 40 km/s respectively \citep{brunetti21}. 

Ultimately, we find that the disruption timescale from tidal shocks can be short enough to affect the age distribution of clusters with $t<1$ Gyr in the central regions of LIRGs. Qualitatively, this difference corresponds to an increase in mass-dependent cluster disruption, where long-term cluster evolution in the central regions of LIRGs are primarily influenced by the external tidal field and ISM density within ongoing mergers. Finally, these results are consistent with a similar UV-optical study presented in \citet{aa20} which looked at the cluster populations for a sample of 6 LIRGs, including NGC 3256. While not focusing on the age distribution slope directly, they note a lack of young lower-mass clusters in the inner region relative to their outer-disk cluster sample for this system.

%Additionally, since the mass and age range covered in each spatial sub-sample is identical we stress that the main driver of these differences cannot be a bias in the sample selection.

\begin{figure}
\centering
     \includegraphics[scale=0.4]{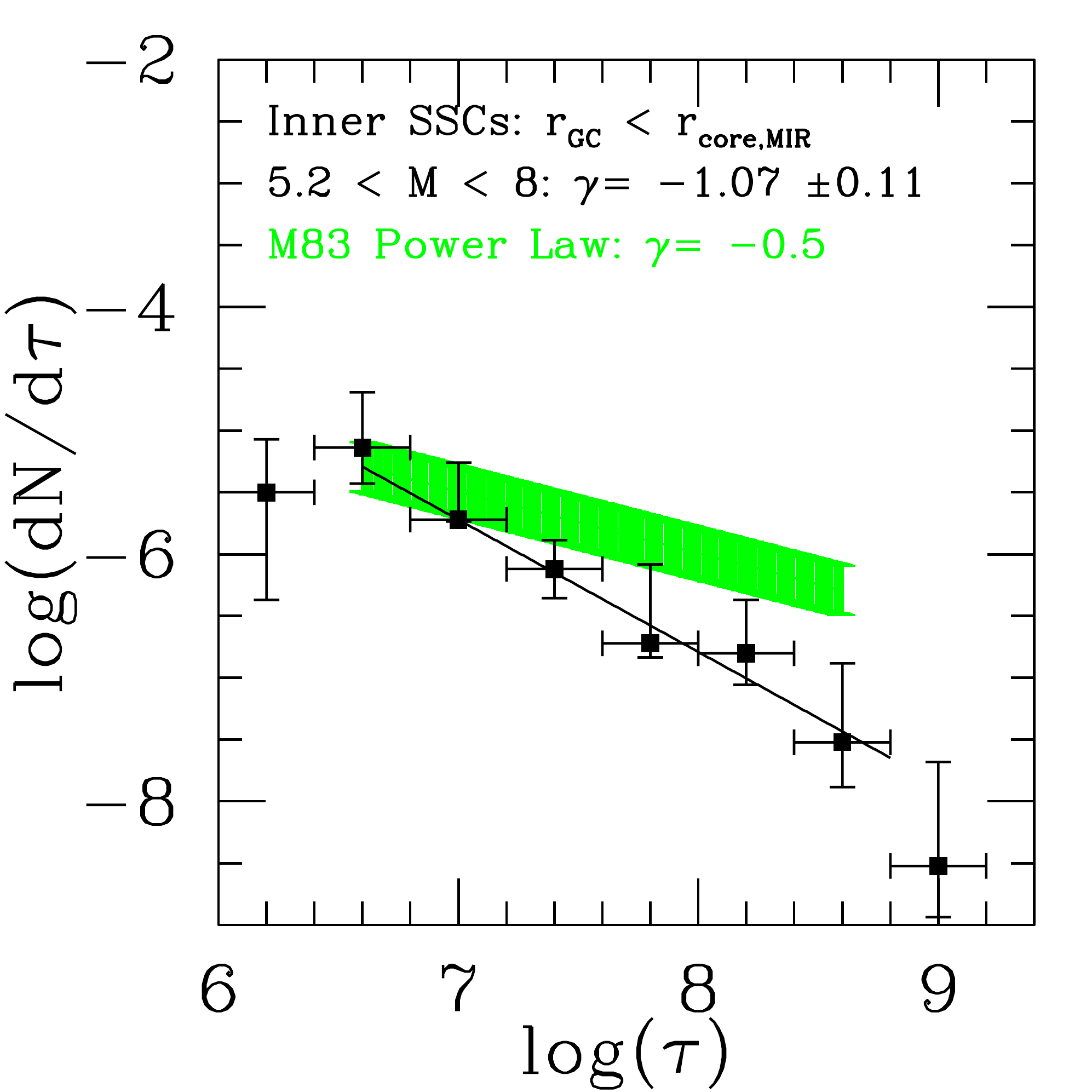}
     \includegraphics[scale=0.4]{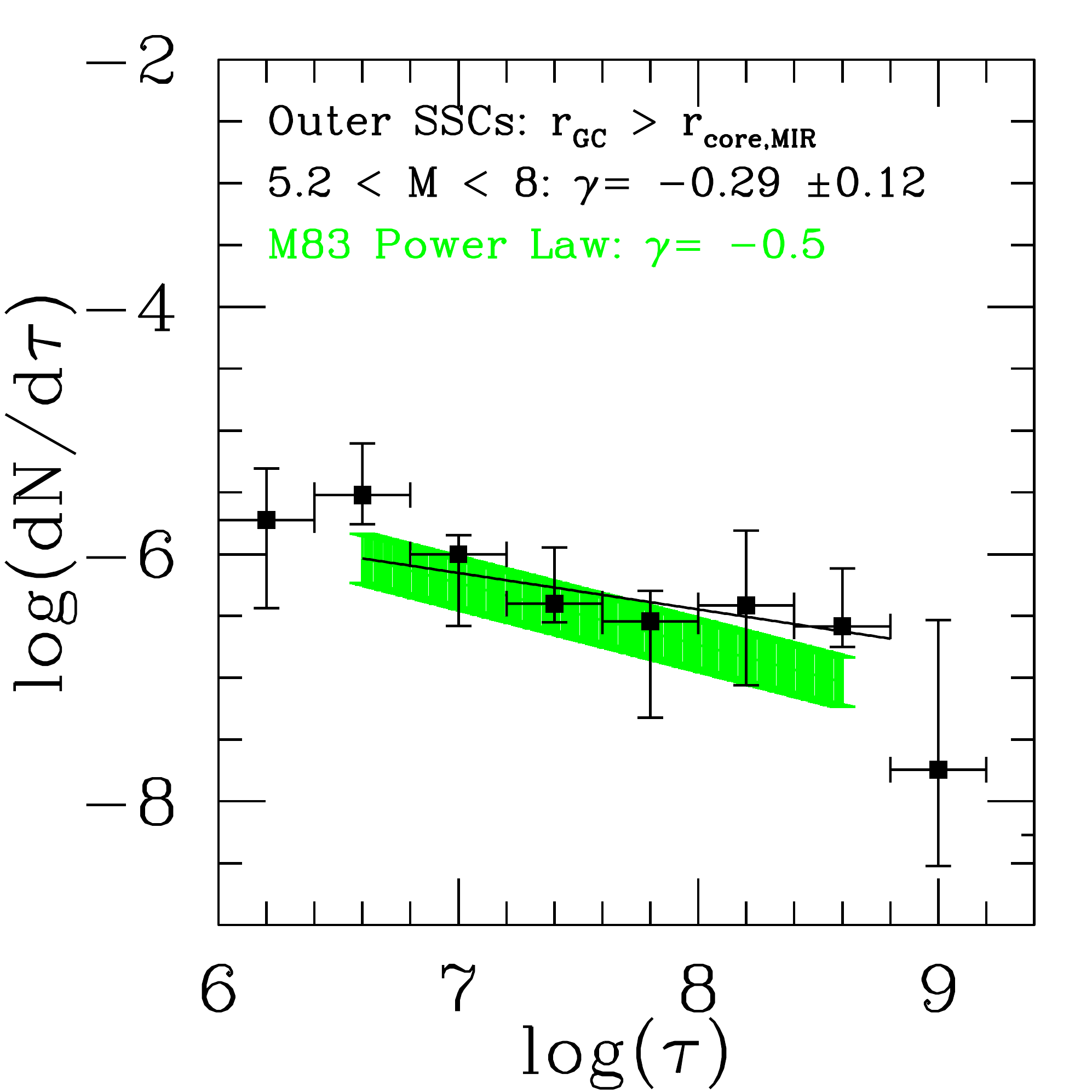}
\caption{The age distribution functions for NGC 3256 and IC 1623 classified as either inner-disk (top) or outer-disk (bottom) clusters relative to the results for the star-forming galaxy M83 shown in green \citep{aab15}. We see that the magnitude of the difference $\gamma$ ($\sim 0.7$)  between the inner- and outer-disk clusters is larger than what has been found for nearby normal galaxies \citep{mm18}, and remains significant even after accounting for spatial variations in the surface density of young cluster formation (See Section 5.1.2). }
\end{figure}

\subsubsection{Increasing SFR vs. Increasing Cluster Destruction}

An important assumption made in the interpretation of these age distributions is that they are the result of cluster disruption, rather than a large increase in the formation of clusters that is a function of distance from the nucleus, or mass. This assumption may not be true in ongoing starbursts like the Antennae or even the central regions of M83. \citet{chandar17} used extinction-corrected GALEX FUV and 24$\mu$m fluxes to derive SFRs averaged from 10 -100 and 100 - 400 My for several starburst galaxies including NGC 3256, M51, and the Antennae. Combined with published color-magnitude diagram (CMD) analysis for the LMC, SMC, NGC 4214, and NGC 4449, they demonstrate that the SFRs have not varied by more than a factor of $\sim2$ for both the dwarf and starburst galaxies in their sample. There also appears to be no systematic trends in the average SFR with age, suggesting that strong variations in the cluster age distribution seen in these systems is likely the result of cluster disruption rather than increased formation. Finally, with complete photometric coverage from the FUV to the FIR \citet{mps15} used the SED fitting code MAGPHYs \citep{magphys08} to derive the SFHs for $\sim 30$ local LIRGs. They find that, apart from Arp 299, galaxies classified as having a "SF burst" show global enhancements of $\sim 2-4$ in SFR, forming $1-2 \%$ of the total stellar mass in the galaxy.

If the differences observed between the inner- and outer-disk age distributions in Figure 8 from is due entirely to an increase in the cluster formation rate (CFR) from 1 - 100 Myr, than the increase in $\gamma$ of 0.7 would correspond to an increase in $\Delta dN/dt \propto CFR$ of 25 for clusters above the completeness limit ($10^{5.2} M_{\odot}$) of our inner-disk sample. While the CFR/SFR relation may vary in different galactic environments, it has been observed to be approximately linear and thus we take the two quantities proportional to one another \citep{lcj17}. The models presented in \citep{dk12b} suggest that the CFR/SFR is $\sim 30-40\%$ for regions with $\Sigma_{gas} \sim 10^{3} M_{\odot}$pc$^{-2}$, and therefore an increase in the CFR of 25 should be considered a lower-limit for the required increase in SFR.

Using the F336W images as a proxy for the recent (10-100 Myr) SF activity we can make estimates of the surface density inside and outside the MIR size of each galaxy. Although the U-band continuum is not tied to the photospheric emission of the youngest stellar populations directly (as is the case for the UV), the errors are minimized for highly star-forming galaxies such as LIRGs \citep{rck98}. We find that the increase in F336W-traced SFR surface density is $6.2 \pm 1.1 M_{\odot}$yr$^{-1}$kpc$^{-2}$ and $3.5 \pm 0.4 M_{\odot}$yr$^{-1}$pc$^{-2}$ for NGC 3256 and IC 1623 respectively. These results are consistent with CALIFA IFU observations of IC 1623 which showed an extinction-corrected enhancement between 30-300 Myr of $\sim 2-3$ in the central half-light radius \citep{cf17}, which would corresponds to a change in $\gamma$ of 0.15-0.2.

Given the above, the most plausible explanation is that while the cluster formation rate in the central regions of NGC 3256 and IC 1623 are clearly increasing, the differences in the slope of the age distribution observed for inner- and outer-disk clusters suggests that cluster disruption plays a key role: clusters found in the inner-regions of these LIRGs show greater disruption rates relative to SSCs identified in their outer-disks. The magnitude of this difference is also larger than the differences observed for inner- and outer-disk clusters in other nearby normal galaxies, suggesting that the merging process is crucial for driving such extreme disruption rates \citep{aa15,hollyhead16}. 

\subsection{Mass Function}

\begin{figure}
\centering
\includegraphics[scale=0.4]{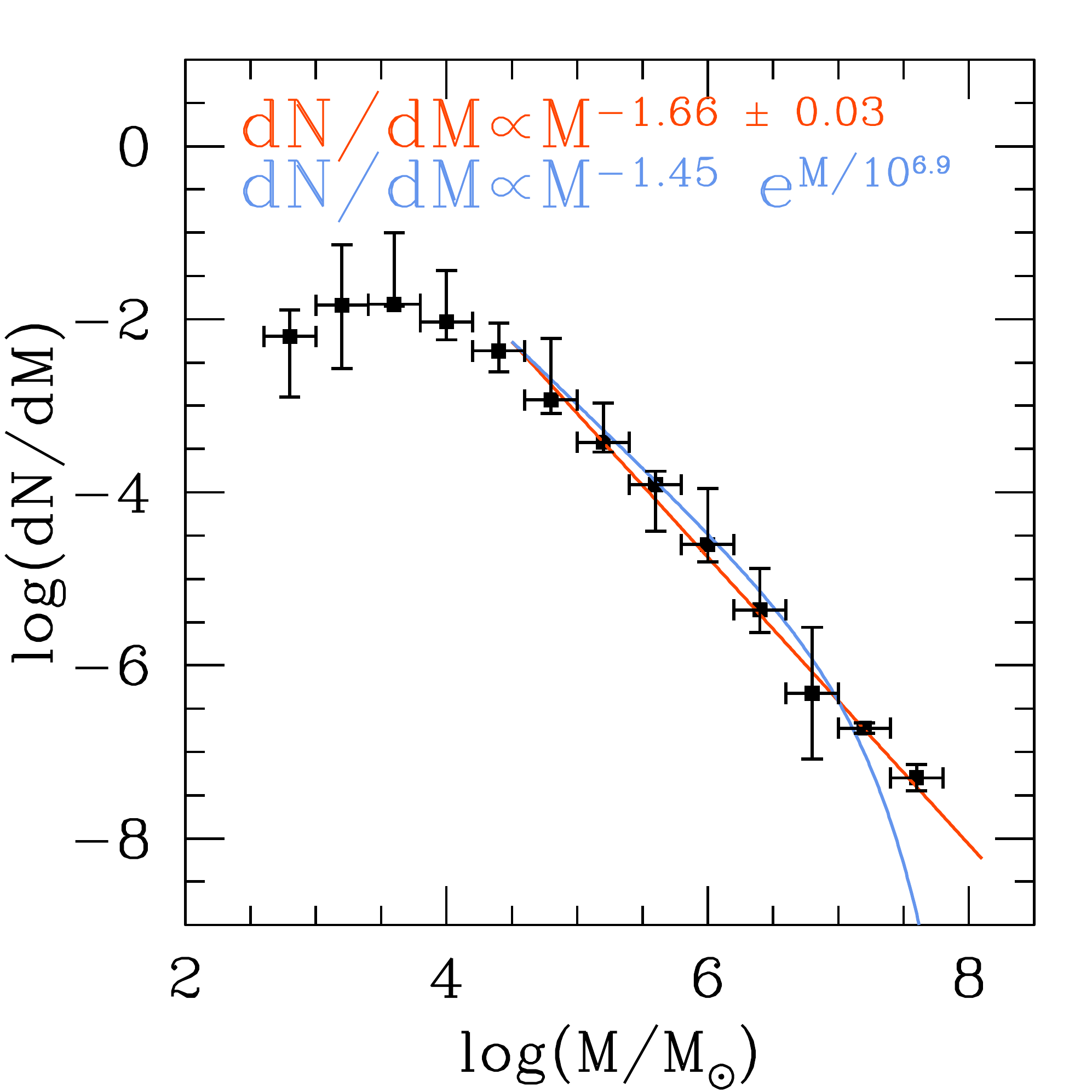}
\caption{The stacked mass distribution function for all 5 galaxies. The red line represents the maximum-likelihood fit to the data for a power-law (PL) only model, with analytic Schechter function (PL+truncation) fit to the empirical distribution overlaid in blue. Although the evidence for a truncation is marginal ($\sim 3\sigma$), the values obtained for the $M_{c}$ are larger than what is observed for lower-luminosity star-forming galaxies in the local Universe. These results are consistent with the luminosity function derived from Pa$\beta$ observations of LIRGs in GOALS \citep{klarson20}, as well as detailed observations of Arp 299 \citep{rand19}.}
\end{figure}

We consider the mass distribution of clusters in our LIRG sample over Regime 2 described in \S 4. When modeling the CMF we adopt a maximum-likelihood fitting technique as well as a bootstrap resampling of the posterior distribution functions for both a simple power-law model (PL) and a two component Schechter function of the form $dN/dM \propto (M/M_{t})^{\alpha} e^{(M/M_{t})}$, where $M_{t}$ is the truncation mass and $\alpha$ is the overall power-law slope. In order to test the hypothesis of a mass truncation, we have fit the cumulative distribution function:

\begin{equation}
 N (M' > M ) = N_{t} \left [(\frac{M}{M_{t}})^{\alpha + 1} - 1\right],
\end{equation}

where $N_{t}$ is the number of clusters more massive than $2^{1/(\alpha + 1)} M_{t}$. We implement the IDL routine MSPECFIT, which has been used previously to study the mass functions of both giant molecular clouds and SSCs in galaxies in the local Universe \citep{rosolowsky07,mm18}. A best-fit value for $N_{t}$ with a $\geq 3\sigma$ significance will imply evidence for a truncation in the mass function at the truncation mass $M_{t}$. In Table 3 we list the best-fit values to the cumulative distribution function for all 5 LIRGs, the 3 early-stage systems, the 2 late-stage systems, and finally the inner- and outer-disk cluster samples which are labeled all, early, late, inner, and outer respectively.

In Figure 9 we show the best-fit PL (red) and PL+truncation (blue) models for the full sample. We find that clusters with ages $t \leq 10^{7.5}$ yr have $\alpha \sim -1.45 \pm 0.03$ and $M_{t} = 10^{6.9} M_{\odot}$. However at such high values of $M_{t}$ our ability to constrain this upper-mass truncation becomes challenging due to the low number of clusters expected above this limit. Indeed, apart from the All and Early-stage distributions, which each have a $N_{t}/\sigma_{N_{t}} \sim 3$ and $M_{t} \geq 10^{6}$, we do not find evidence supporting a truncated power-law in any of the remaining sub-samples. \citet{aa20} find evidence, albeit for 1/6 targets that the inclusion of older clusters with ages up to 100 Myr, can increase the significance of the PL+truncation model to $\geq 3\sigma$. However we do not find that including older clusters in any of our sub-samples increases the significance of the PL+truncation model such that we favor it over the simple PL model.

Examining the results for inner- and outer-disk clusters we see that the inner-disk clusters drive the trend seen in the larger dataset in support of a truncated power-law model. Given that the inner-disk has been shown to be a highly-disruptive environment for young clusters, it is not unexpected to find stronger evidence of a truncation. Overall, the PL models for the inner- and outer-disk clusters agree within uncertainties. This is consistent with detailed results for another well-studied LIRG, Arp 299, which show differences in the derived $M_{t}$ of $\leq$ a factor of 2 between the two galaxy nuclei and the extended disk \citep{rand19}.

\input{tbl-3}

Finally, our value for the slope of the CMF is consistent with the slope of the luminosity functions derived from Pa$\beta$ observations of LIRGs in GOALS \citet{klarson20}. By comparison, $\alpha$ is commonly measured to be $-2$ for lower luminosity star-forming galaxies with an $M_{t} \sim 10^{3-4} M_{\odot}$ \citep{nb08b,larsen10}. These results suggest that LIRGs are capable of forming relatively more massive clusters that what is seen in other nearby galaxies.

%Further, recent hydrodynamical simulations of cluster formation in nearby starburst galaxies reveal that the underlying CMF is steeper when the observed $M_{c}$ increases. Thus, due to the limited sensitivity of the observations in \citet{stl17}, we expect to observe a shallower CMF for these new observations which span a much larger range in cluster mass, and which detect clusters at $M \sim 10^{4} M_{\odot}$ \citep{maji17, lahen20}.
%Further, by limiting our analysis of the CMF to clusters with $t < 10^{7.5}$yr, spatial differences in the observed cluster disruption will not be strongly reflected in the underlying CMF. It is only by examining older clusters, which can not be done effectively for inner-disk clusters due to completeness, that the differences in the disruption rate would result in significant changes to the observed CMF \citep{dk12a}.

\subsubsection{Merger Stage Dependence}

\begin{figure}
\centering
     \includegraphics[scale=0.4]{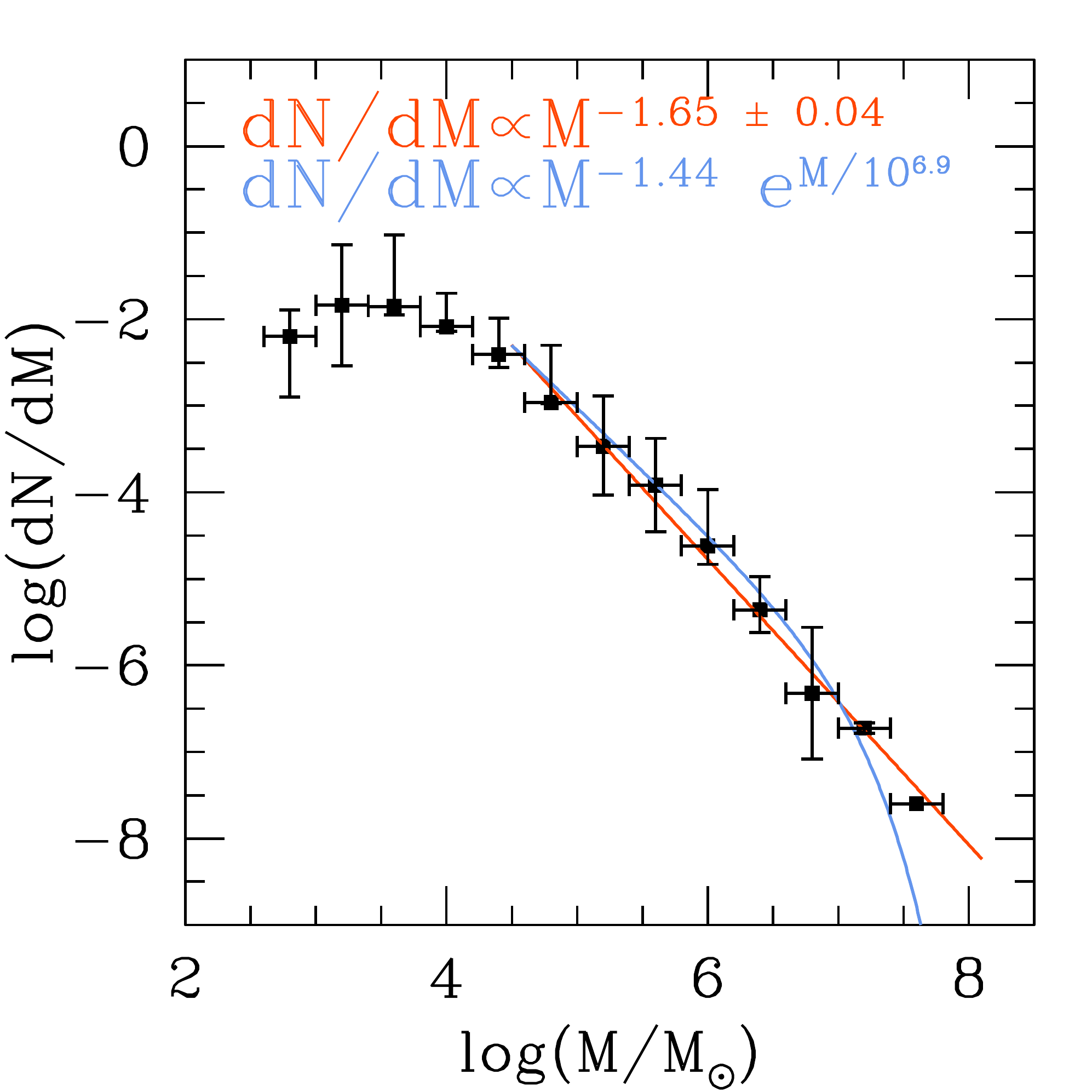}
     \includegraphics[scale=0.4]{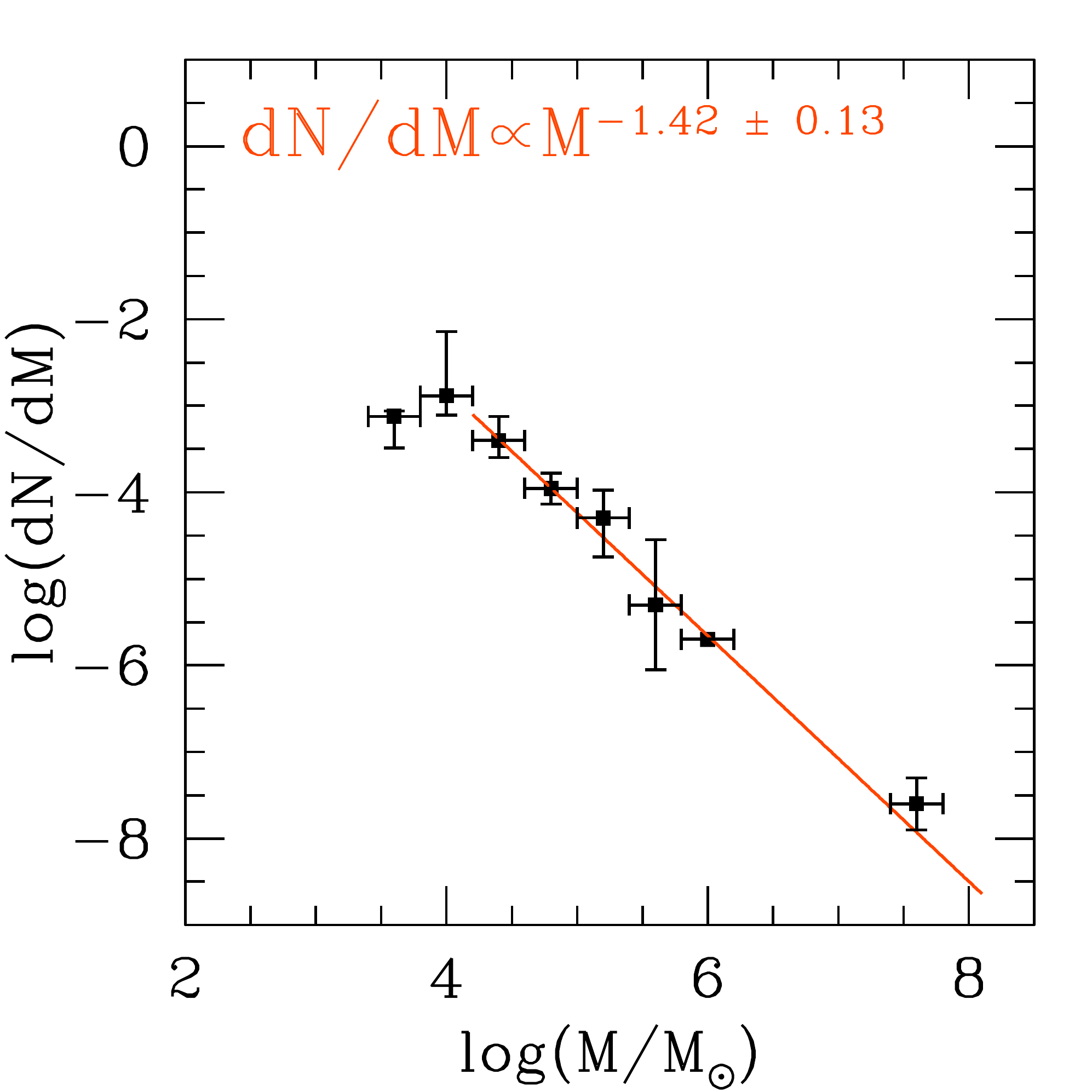}
\caption{The mass distribution function for early- (top Panel) and late-stage (bottom Panel) mergers showing a flattening of the observed CMF. This result is consistent with the idea that the progressive destruction of lower-mass clusters (depending on the exact value of $\delta$) flattens the observed CMF over time such that the late-stage mergers show a statistically flatter slope.}
\end{figure}

Since U/LIRGs in the GOALS sample span the full range of merger stages, we can test if our explanation of cluster formation and destruction depends on the dynamical state of the galaxy. \citet{haan13}, \citet{dck13}, and \citet{ss13a} have classified the merger stage of each U/LIRG in the GOALS sample based on their morphological appearance at multiple wavelengths. These merger classification schemes run from pre-first passage to single coalesced nuclei. In order to search for changes in the cluster age and mass distributions as a function of merger stage, we separated our sample into early (classes 0-3: NGC 3256, IC 1623, NGC 7469) and late-stage (classes 4-6: Arp 220, NGC 6240) mergers. 
s
In Figure 10 we show the CMF for Regime 2 for both the early- and late-stage mergers and find that young star clusters in the early-stage show a power-law distribution of $dN/dM \propto M^{-1.65 \pm 0.04}$ consistent with the slope for the full sample. However, when examining the slope for the late-stage mergers we find that it appears flatter ($\alpha = -1.42 \pm 0.13$). For the late-stage mergers we extend Regime 2 down to $10^{4.2} M_{\odot}$ given that these systems are not used in the inner- and outer-disk comparison, and that many of the clusters live \textit{outside} the MIR size defined for these systems. This result suggests that late-stage mergers may be able to form massive clusters with an even higher efficiency (increased proportion of higher-mass clusters formed) than what has been observed for LIRGs involved in early-stage mergers. Indeed, the two late-stage mergers are both ULIRGs with a median SFR that is $\sim 2$x higher than the early-stage mergers in our sample.

In order to quantify any differences in the age distributions for each merger stage we ran a KS-test comparing the normalized distributions of the early- and late-stage mergers to the total sample (see Figure 11). We find that within our sub-sample of 5 LIRGs there is a $\sim 3 \sigma$ difference in the age distribution of clusters between our early- and late-stage mergers, with the latter showing a much larger fraction of old massive clusters. As we have shown in Figure 8, these clusters are more likely to survive in the outer-disks of early-mergers, where the effect of tidal shocks is less severe compared to the inner-disks. Hence, the fact that we see a higher fraction of older clusters in late-stage mergers, where most clusters are found in the outer-disks, can be directly attributed to the higher survival rates of older clusters in the outer-disks that allow them to accumulate over the course of the merger.

Given the above, the data supports a scenario where the flat CMF for young clusters is due primarily to an increase in the formation of young massive clusters, which results in a flattened mass distribution in Regime 2 relative to the early-stage mergers. However there is a significant population of old massive clusters which have survived the progressive increase in the destruction of lower-mass clusters (depending on the exact value of $\delta$) throughout the course of the merger. Ultimately, higher SFRs, stronger tidal shocks, and increasing pressures throughout a galaxy merger all play a role in shaping the evolution of the cluster age and mass distributions relative to nearby low-luminosity star-forming galaxies.

\begin{figure}
\centering
\includegraphics[scale=0.4]{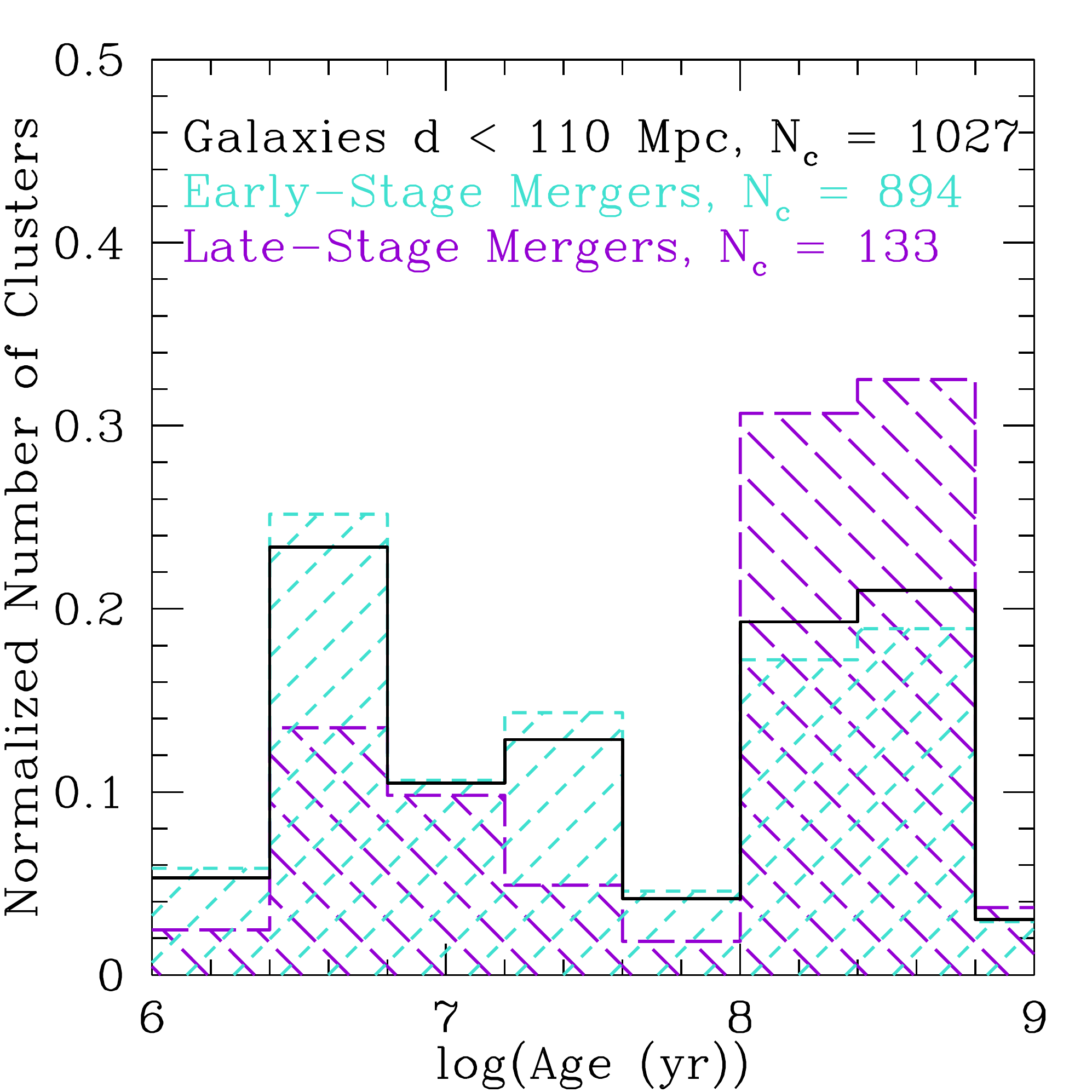}
\caption{The binned cluster ages for early- and late-stage LIRGs normalized to the total number of clusters in each respective sub-sample. A KS-test reveals a $3\sigma$ difference between the distribution of early- and late-stage cluster ages. This suggests that a significant evolution has taken place within the cluster population of the late-stage mergers.}
\end{figure}

%Again I'm not sure about what you try to mean here. 

%BEGIN SUMMARY SECTION~~~~~~~~~~~~~~~~~~~~~~~~~~~~~~~~~~~~~~~~
\section{Summary}

{\it Hubble Space Telescope} WFC3/UVIS (F336W) and ACS/WFC optical (F435W and F814W) observations for a sample of 10 LIRGs in the GOALS sample were obtained. These observations have been used to derive the ages, masses, and extinctions of the super star clusters contained within the 5 nearest systems in order to examine the cluster properties in extreme starburst environments relative to those in nearby, lower luminosity star-forming galaxies. The following conclusions are reached:

\noindent
(1) We have detected 1027 clusters throughout the disks of these 5 LIRGs. These clusters have $S/N \geq 3$ in all three filters and de-convolved FWHMs as measured by ISHAPE of $\leq 3$ pixels, corresponding to a median physical size of $\sim$22pc.

\noindent
(2) The derived cluster age distribution implies a disruption rate of $dN/d\tau \propto \tau^{-0.5 +/- 0.2}$ over $10^{6.2} < \tau < 10^{8.7}$ and for cluster masses $\geq 10^{5.2} M_{\odot}$. This is consistent with the general framework that ongoing mergers destroy clusters at a much higher rate due to the high external pressure induced from a strong tidal field. The measured $\gamma$ is steeper than what has been observed in lower-luminosity star-forming galaxies in the local Universe.

\noindent
(3) Differences in the slope of the observed cluster age distribution between inner- and outer-disk clusters ($\Delta \gamma \sim 0.7$) provides some of the clearest evidence to date of location-dependent cluster destruction, and that this destruction outweighs the increase in the CFR within the central region. Not only does this effect appear to be common in LIRGs, but the magnitude of this affect is amplified even relative to strong gradients of cluster disruption observed in normal star-forming galaxies in the local Universe. 

\noindent
(4) The derived cluster masses imply a CMF for our sample galaxies of $dN/dM = M^{-1.66 +/- 0.03}$ with marginal evidence for a PL+truncation at $M_{t} \sim 2$x$10^{6} M_{\odot}$ for early-stage mergers, which dominate the full sample. This trend appears to be driven in large part by the inner-disk, where the extreme environment acts to form massive clusters at a rate which exceeds expectations from a $-2$ power-law at the highest masses.

\acknowledgements
The authors thank Matteo Messa for useful discussions regarding spatially-variable cluster completeness and assistance in running the LEGUS Cluster Completeness Tool on our dataset. 

S.T.L was partially supported thorough NASA grant HST-GO15472. A.S.E. was supported by NSF grant AST 1816838 and by NASA through grants HST-GO10592.01-A, HST-GO11196.01-A, and HST-GO13364 from the Space Telescope Science Institute, which is operated by the Association of Universities for Research in Astronomy, Inc., under NASA contract NAS5- 26555. V.U acknowledges funding support from NASA Astrophysics Data Analysis Program (ADAP) Grant 80NSSC20K0450. HI acknowledges support from JSPS KAKENHI Grant Number JP21H01129. Finally, This research has made use of the NASA/IPAC Extragalactic Database (NED) which is operated by the Jet Propulsion Laboratory, California Institute of Technology, under contract with the National Aeronautics and Space Administration.

\bibliography{master_ref}

\appendix

\input{mastercat_der_sb99_round}
\input{mastercat_obs_sb99_round}

\begin{figure*}
  \centering
  \includegraphics[scale=0.4]{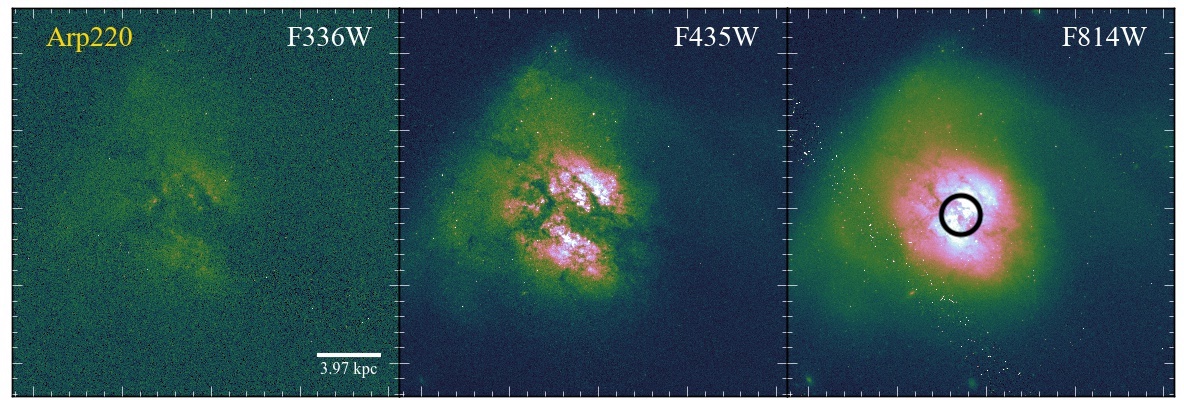}
   \caption{Same as Figure 1}
\end{figure*}

\begin{figure*}
  \centering
  \includegraphics[scale=0.4]{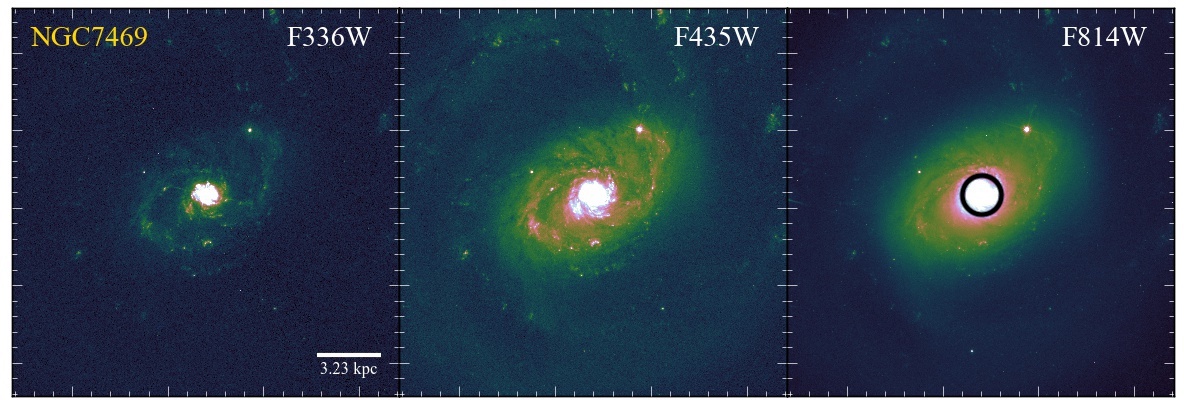}
   \caption{Same as Figure 1}
\end{figure*}

\begin{figure*}
  \centering
  \includegraphics[scale=0.4]{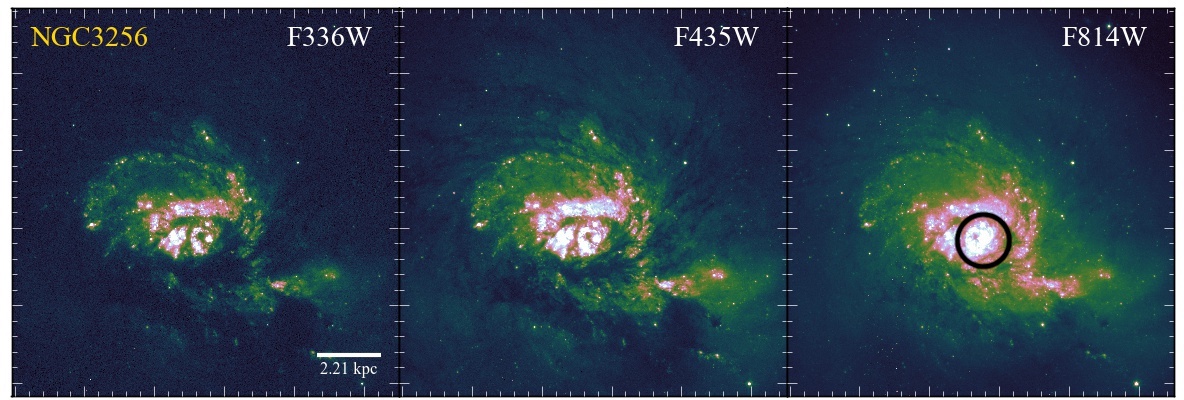}
  \caption{Same as Figure 1}
\end{figure*}

\begin{figure*}
  \centering
  \includegraphics[scale=0.4]{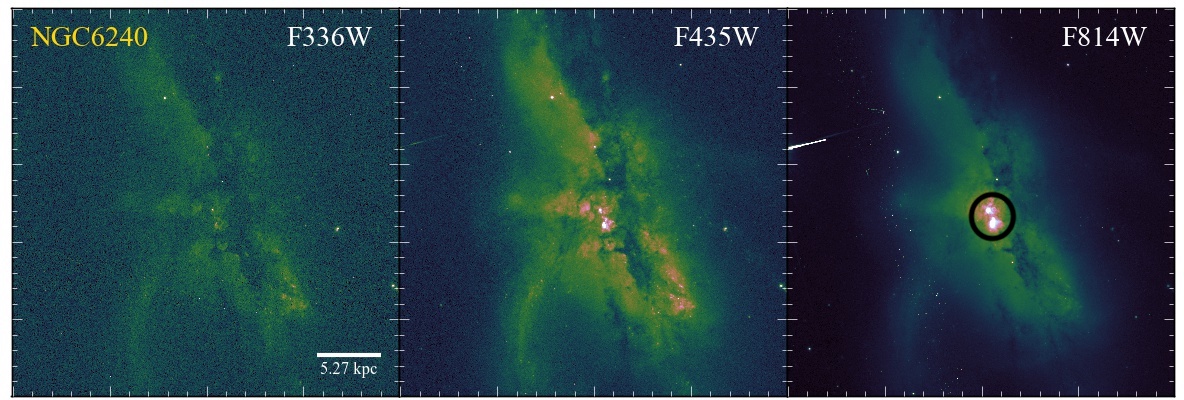}
  \caption{Same as Figure 1}
\end{figure*}

\begin{figure*}
  \centering
  \includegraphics[scale=0.4]{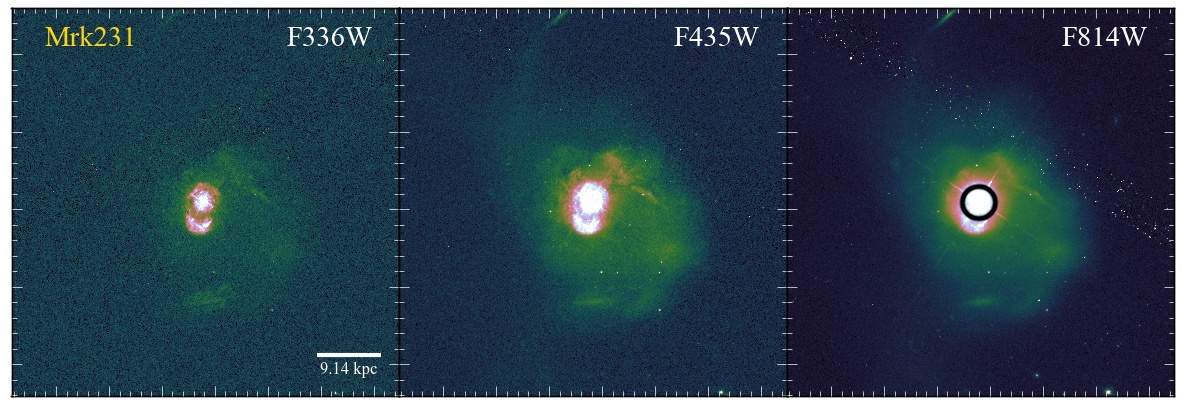}
  \caption{Same as Figure 1}
\end{figure*}

\begin{figure*}
  \centering
  \includegraphics[scale=0.4]{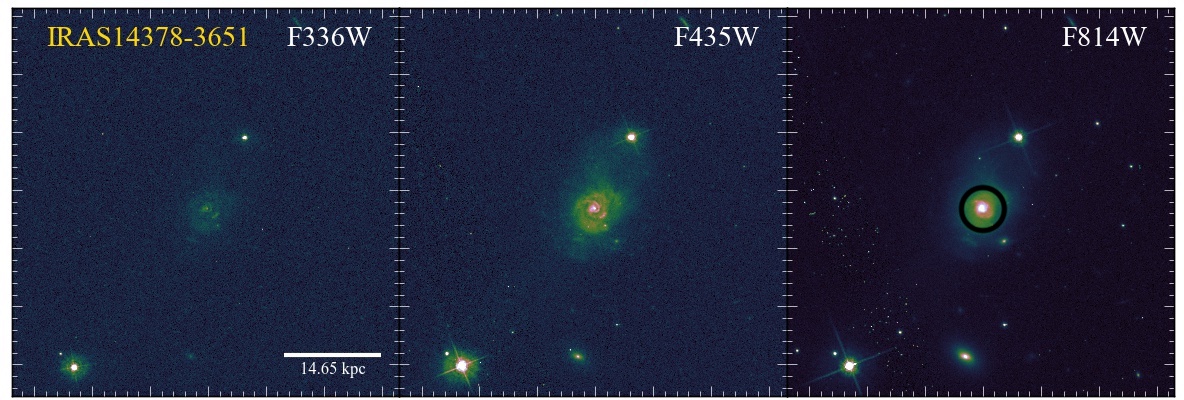}
  \caption{Same as Figure 1}
\end{figure*}

\begin{figure*}
  \centering
  \includegraphics[scale=0.4]{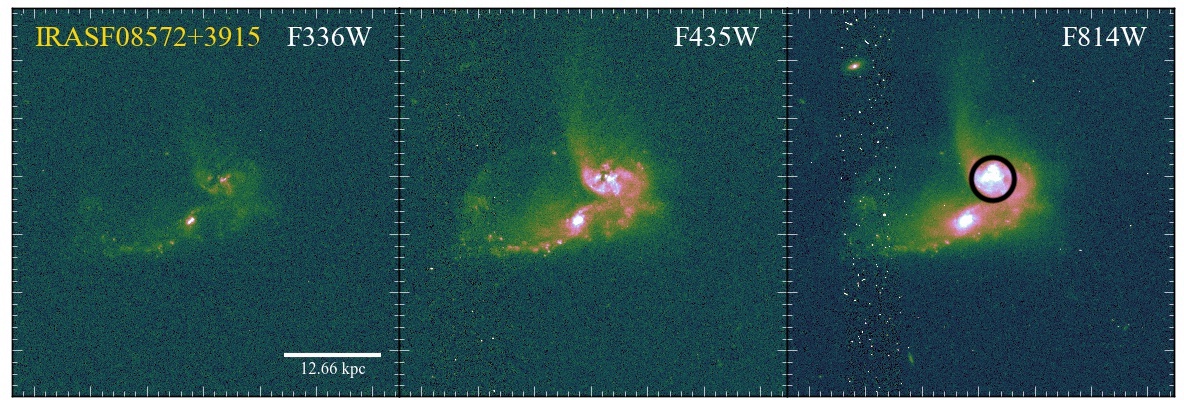}
  \caption{Same as Figure 1}
\end{figure*}

\begin{figure*}
  \centering
  \includegraphics[scale=0.4]{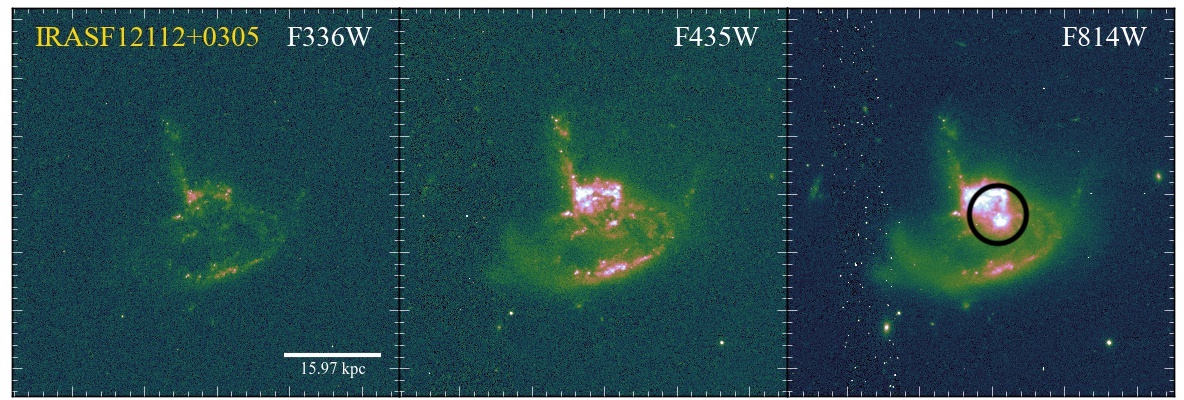}
  \caption{Same as Figure 1}
\end{figure*}

\begin{figure*}
  \centering
  \includegraphics[scale=0.4]{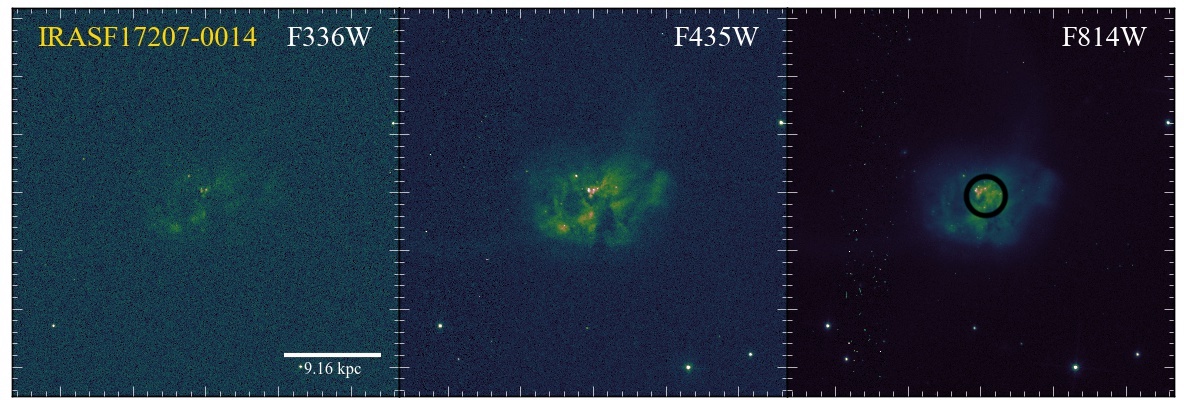}
  \caption{Same as Figure 1}
\end{figure*}

\begin{figure*}
\centering
\includegraphics[scale=0.6]{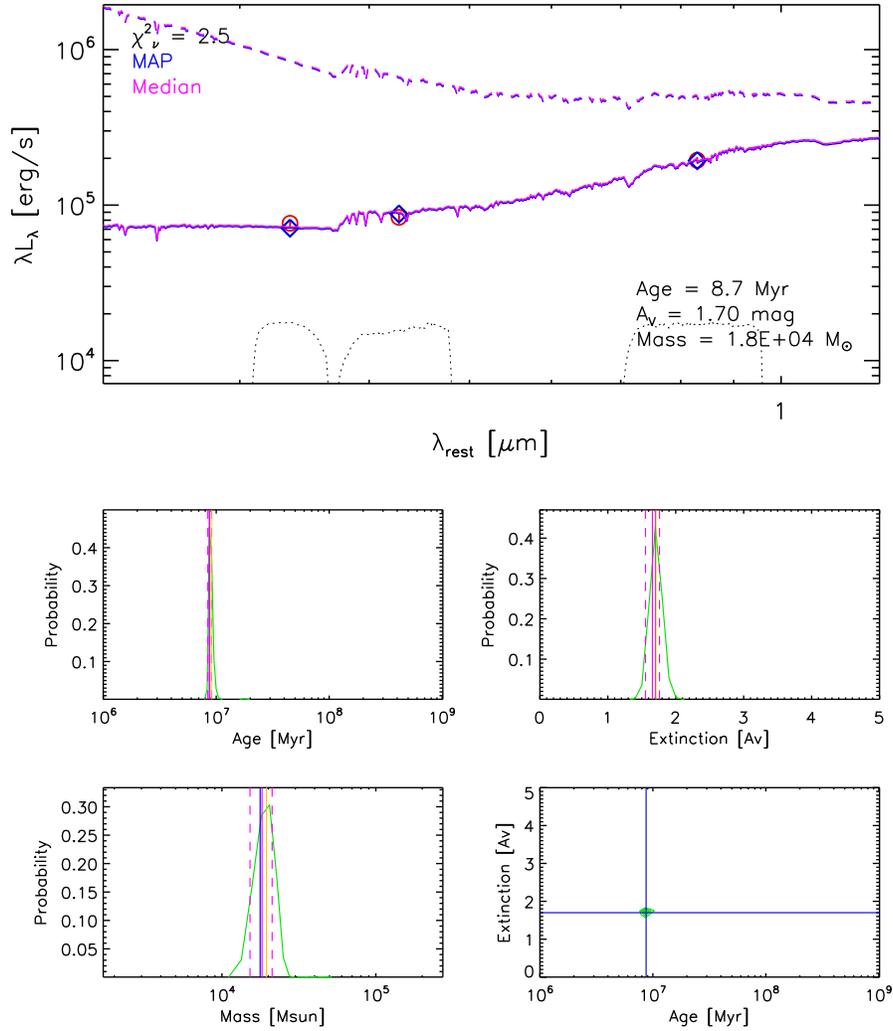}
\caption{Example SED and resulting model fit for cluster 33 in NGC 3256 with an age of 8.7 Myr (blue) with an extinction of $A_V  = 1.7$ mags. The dashed lines correspond to the un-extincted SEDs for each model. In the example SEDs provided, this combination of filters can be used to uniquely solve for cluster age relative to the F140LP observations used in our previous study. Finally, in the lower Panels the resulting $1\sigma$ errors for the age, mass, and extinction are shown as dashed pink lines superimposed over the marginalized posterior distribution for each parameter (green curves).}
\end{figure*}

\end{document}

%% file: tbl-1_full_v2.tex
\begin{deluxetable*}{l|llcccccccc}
\center
\tablecaption{Galaxy Sample Properties \label{tbl-1}}
\tabletypesize{\footnotesize}
\tablewidth{0pt}
\tablehead{
\colhead{Galaxy}  & \colhead{RA} & \colhead{Dec} & \colhead{$D_{L}$ (Mpc)}  & \colhead{MS} & \colhead{Log(LIR)} & \colhead{Log($M_{*}$)} &  \colhead{SFR ($M_{\odot}$/yr)} & \colhead{MIR Size (kpc)} & \colhead{b/a (deg)} & \colhead{PA (deg)}}
\startdata
IC 1623 &  01h07m47.180s & -17d30m25.30s & 84.40 & 3 & 11.71 & 11.21 & 94.09 & 4.60 & 0.92 & 302.7 \\
IRAS F08572+3915 &  09h00m25.390s & +39d03m54.40s & 261.13 & 3 & 12.16 & 11.81 & 254.32 & 4.70$^{*}$ & 0.31 & 10.8 \\
NGC 3256 & 10h27m51.270s & -43d54m13.49s & 45.66 & 3 & 11.64 & 11.06 & 76.46 & 2.10 & 0.83 & 25.0 \\
IRAS F12112+0305 &  12h13m46.00s & +02d48m38.0s & 329.38 & 3 & 12.36 & 11.34 & 402.89 & 9.83 & 0.64 & 9.3  \\
Mrk 231$^{a}$ &  12h56m14.234s & +56d52m25.24s & 188.57 & 6 & 12.50 & 10.98 & 380.19 & 3.34$^{*}$ & 0.83 & 25.0\\
IRAS 14378-3651 &  14h40m59.008s & -37d04m31.94s & 302.13 & 6 & 12.23 & --- & 79.0$^{b}$  & 5.78$^{*}$ & 0.67 & 41.7\\
Arp 220 &  15h34m57.255s & +23d30m11.30s & 81.93 & 5 & 12.28 & 11.06 & 327.74 & 1.53$^{*}$ & 0.87 & 286.2\\
NGC 6240 & 16h52m58.871s & +02d24m03.33s & 108.67 & 4 & 11.93 & 11.59 & 148.44 & 2.26 & 0.45 & 279.6\\
IRAS F17207-0014 &  17h23m21.955s & -00d17m00.94s & 189.03 & 5 & 12.46 & 11.18 & 501.22 & 3.75$^{*}$ & 0.80 & 30.7\\
NGC 7469 & 23h03m15.623s & +08d52m26.39s & 66.68 & 2 & 11.65 & 11.38 & 80.26 & 1.37 & 0.53 & 56.3\\
\enddata
\tablecomments{Infrared luminosity and stellar mass are given in units of $L/L_{\odot}$ and $M/M_{\odot}$ respectively. RA, Dec, distance, infrared luminosity, stellar mass, and SFR are taken from \citet{jh10}. Merger stages (MS) are taken from \citet{haan13}, \citet{dck13}, and \citet{ss13a}.  MIR sizes are taken from \citet{tds10}. The values marked with an asterisk ($*$) denotes a galaxy whose central region is unresolved in the Spitzer 2D IRS Spectra. The b/a and PA are taken from \citet{dck13}.}
\tablenotetext{a}{Data for Mrk 231 is taken from \citet{vu12}}
\tablenotetext{b}{An estimate for the SFR of IRAS 14378-3651 is taken from \citet{sturm11}}
\end{deluxetable*}

%% file: tbl-2.tex
\begin{deluxetable*}{l|cccccccccc}
\center
\tablecaption{Cluster Detection Statistics and Overall Properties \label{tbl-2}}
\tabletypesize{\footnotesize}
\tablewidth{0pt}
\tablehead{
\colhead{Galaxy}  & \colhead{$N_{det}$} & \colhead{$N_{c}$} & \colhead{m$_{U, Outer} 90\%$} & \colhead{m$_{U, Inner} 90\%$}  & \colhead{$t_{c}$} & \colhead{$\sigma_{t_{c}}$} & \colhead{$M_{c}$} & \colhead{$\sigma_{M_{c}}$} &  \colhead{$A_{V}$} & \colhead{$\sigma_{A_{V}}$}}
\startdata
Arp 220 & 108 & 46 & 24.19 & 23.28 & 8.33 & 0.53 & 5.94 & 0.91 & 0.7 & 0.48 \\
IC 1623 & 453 & 180 & 24.85 & 23.84 & 6.97 & 0.53 & 5.21 & 0.89 & 0.4 & 0.58 \\
NGC 3256 & 1082 & 549 & 24.51 & 23.82 & 7.99 & 1.06 & 5.29 & 1.17 & 0.6 & 0.75 \\
NGC 6240 & 141 & 87 & 24.95 & 23.36 & 8.24 & 0.66 & 5.71 & 1.16 & 0.5 & 0.59 \\
NGC 7469 & 383 & 165 & 24.41 & 23.71 & 7.24 & 0.79 & 4.82 & 0.95 & 0.4 & 0.55 \\
\enddata
\tablecomments{$N_{det}$ is the number of cluster candidates in each galaxy. $N_{c}$ is the number of confirmed clusters. m$_{U} 90\%$ for both the inner and outer regions in each galaxy are listed in apparent magnitudes. Values for cluster age, mass, and extinction are given in log(t/yr), $M/M_{\odot}$, and magnitudes respectively. We note that a $3\sigma$ significance for the PL+truncation model $N_{c}/\sigma_{N_{c}}$ is only reached for the All and Early-stage cluster sub-samples.}
\end{deluxetable*}

%% file: tbl-3.tex
\begin{deluxetable*}{l|cccccc}
\center
\tablecaption{Results From MSPECFIT for PL and PL+Truncation \label{tbl-3}}
\tabletypesize{\footnotesize}
\tablewidth{0pt}
\tablehead{
\colhead{Sample}  & \colhead{$N_{t}$} & \colhead{$\sigma_{N_{t}}$} & \colhead{$M_{t}$} & \colhead{$\sigma_{M_{t}}$}  & \colhead{$\alpha$} & \colhead{$\sigma_{\alpha}$}}
\startdata
All & 24.22 & 7.43 & 8.15e+06 & 5.74e+06 & -1.45 & 0.03 \\ 
All-PL & - & - & - & - & -1.66 & 0.03 \\ 
Early & 23.73 & 7.61 &  8.52e+06 &  2.51e+06 & -1.45 & 0.03 \\ 
Early-PL & - & - & - & - & -1.65 & 0.02 \\
Late & 0.491 & 2.02 & 6.62e+07 & 1.68e+07 & -1.42 & 0.09 \\
Late-PL & - & - & - & - & -1.42 & 0.13 \\
Inner & 32.83 &	 13.3 & 6.38e+05 & 1.71e+06 & -1.51 & 0.17 \\
Inner-PL & - & - & - & - & -1.58 & 0.08 \\
Outer & 3.63 & 2.65 & 1.32e+07 & 1.18e+07 & -1.60 & 0.06 \\
Outer-PL & - & - & - & - & -1.65 & 0.06 \\
\enddata
\tablecomments{$N_{t}$ is the number of clusters more massive than $2^{1/(\alpha + 1)} M_{t}$, where $M_{t}$ is the truncation mass in the Schechter function given in $M/M_{\odot}$. $\alpha$ is the slope of the PL or the PL component of the Schchter function. We note that a $3\sigma$ significance for the PL+truncation model $N_{t}/\sigma_{N_{t}}$ is only reached for the All and Early-stage cluster sub-samples.}
\end{deluxetable*}

%% file: mastercat_der_sb99_round.tex
\startlongtable
\begin{deluxetable*}{lllcccccc}
\tablecolumns{9}
\tablecaption{Derived Properties of Individual Star Clusters \label{tbl-5}}
\tabletypesize{\footnotesize}
\tablewidth{0pt}
\tablehead{
\colhead{ID} & \colhead{R.A.} & \colhead{Decl.} & \colhead{$t_{c}$} & \colhead{$\sigma_{t_{c}}$} & \colhead{$M_{c}$} & \colhead{$\sigma_{M_{c}}$} & \colhead{$A_{V}$} & \colhead{$\sigma_{A_{V}}$}
}
\startdata
\cutinhead{Arp 220}
1  & 233.746669054 & 23.480581701 & 8.73 & 0.13 & 9.02 & 0.05 & 4.50 & 0.22 \\
2  & 233.729549053 & 23.475817413 & 8.02 & 0.38 & 7.07 & 0.19 & 0.80 & 0.31 \\
3  & 233.748437605 & 23.482745482 & 9.00 & 0.42 & 6.49 & 0.11 & 0.90 & 0.65 \\
4  & 233.751644007 & 23.484386024 & 8.26 & 0.09 & 6.65 & 0.05 & 0.60 & 0.13 \\
5  & 233.764694288 & 23.492547085 & 8.73 & 0.15 & 7.83 & 0.05 & 3.40 & 0.24 \\
6  & 233.728218353 & 23.482656154 & 6.48 & 0.35 & 5.16 & 0.37 & 3.00 & 1.02 \\
7  & 233.732302621 & 23.484575842 & 6.97 & 0.02 & 5.32 & 0.05 & 2.10 & 0.10 \\
8  & 233.731859917 & 23.486152064 & 8.26 & 0.27 & 5.49 & 0.14 & 0.80 & 0.23 \\
\enddata
\tablecomments{All values for age, mass, and extinction are given in Log(t/yr), Log($M/M_{\odot}$), and magnitudes respectively}
\end{deluxetable*}

%% file: mastercat_obs_sb99_round.tex
\startlongtable
\begin{deluxetable*}{lllcccccc}
\tablecolumns{9}
\tablecaption{Observational Properties of Individual Star Clusters \label{tbl-4}}
\tabletypesize{\footnotesize}
\tablewidth{0pt}
\tablehead{
\colhead{ID} & \colhead{R.A.} & \colhead{Decl.} & \colhead{$f_{336}$} & \colhead{$\sigma_{f_{336}}$} & \colhead{$f_{435}$} & \colhead{$\sigma_{f_{435}}$} & \colhead{$f_{814}$} & \colhead{$\sigma_{f_{814}}$}
}
\startdata
\cutinhead{Arp 220}
1  & 233.746669054 & 23.480581701 & 1.07E-18 & 3.20E-20 & 9.25E-18 & 1.51E-20 & 5.17E-17 & 7.53E-20 \\
2  & 233.729549053 & 23.475817413 & 1.70E-17 & 9.12E-20 & 2.64E-17 & 1.95E-20 & 1.25E-17 & 2.07E-20 \\
3  & 233.748437605 & 23.482745482 & 2.94E-19 & 2.42E-20 & 8.17E-19 & 9.83E-21 & 7.72E-19 & 4.42E-21 \\
4  & 233.751644007 & 23.484386024 & 4.57E-18 & 5.31E-20 & 8.37E-18 & 1.33E-20 & 3.47E-18 & 7.76E-21 \\
5  & 233.764694288 & 23.492547085 & 3.30E-19 & 2.67E-20 & 2.08E-18 & 1.10E-20 & 6.02E-18 & 9.32E-21 \\
6  & 233.728218353 & 23.482656154 & 5.77E-19 & 3.30E-20 & 6.28E-19 & 1.44E-20 & 4.25E-19 & 9.87E-21 \\
7  & 233.732302621 & 23.484575842 & 5.76E-19 & 2.97E-20 & 4.69E-19 & 1.18E-20 & 1.21E-18 & 7.63E-21 \\
8  & 233.731859917 & 23.486152064 & 2.39E-19 & 2.96E-20 & 4.50E-19 & 1.32E-20 & 2.18E-19 & 7.02E-21 \\
\enddata
\tablecomments{All flux values and $1\sigma$ uncertainties are given in ergs/s/cm$^{2}$/\AA}
\end{deluxetable*}